\documentclass[leqno]{article}
\usepackage[english]{babel}
\usepackage[ansinew]{inputenc}
\usepackage{amsthm}
\usepackage{amscd}
\usepackage{amssymb}
\usepackage{amsmath}
\usepackage{verbatim}
\usepackage{graphics}
\usepackage{graphicx}
\usepackage{color}
\usepackage{multirow}

\newtheorem*{myack}{Acknowledgements}
\newtheorem{myexa}{Example}

\newtheorem{mytheo}{Theorem}
\newtheorem{mylem}{Lemma}
\newtheorem{myprop}{Proposition}
\newtheorem*{myrema}{{\em Remark}}

\newtheorem*{myproo}{{\em Proof}}

\newcommand{\Cov}{\,\mbox{{\rm Cov}}\,}
\newcommand{\Var}{\,\mbox{{\rm Var}}\,}

\definecolor{aggrey}{rgb}{.7,.7,.7}

\begin{document}
\thispagestyle{empty} \vspace*{-1cm}

\begin{center}
{\sc  Minimum Risk Equivariant Estimation of the Parameters of the}\\
{\sc General Half-Normal Distribution by Means of a }\\
{\sc Monte Carlo Method to Approximate Conditional Expectations}\vspace{1cm}\\
A.G. Nogales, P. P\'erez and P. Monfort\vspace{1cm}\\
Dpto. de Matem\'aticas, Universidad de Extremadura\\
Avda. de Elvas, s/n, 06071--Badajoz, SPAIN.\\
e-mail: nogales@unex.es
\end{center}
\vspace{.4cm}
\begin{quote}
\hspace{\parindent} { {\sc Abstract.} This work addresses the
problem of estimating the parameters of the general half-normal
distribution. Namely, the problem of determining the minimum risk
equi\-va\-riant (MRE) estimators of the parameters is explored. 
Simulation studies are realized to compare the behavior of these
estimators with maximum likelihood and unbiased estimators. A natural Monte Carlo method to compute conditional expectations is used to approximate the MRE estimation of the location parameter because its expression involves two conditional expectations not easily computables. The used Monte Carlo method is justified by a theorem of Besicovitch on differentiation of measures, and has been slightly modified to solve a sort of ``curse of dimensionality" problem appearing in the estimation of this parameter. This method has been implicitly used in the last years in the context of ABC (approximate Bayesian computation) methods.}
 \end{quote}

\vfill
\begin{itemize}
\item[] \hspace*{-1cm} {\em AMS Subject Class.} (2010): {\em Primary\/}
62B05 {\em Secondary\/} 62F10.
\item[] \hspace*{-1cm} {\em Key words and phrases:} General half-normal distribution, equivariance, Monte Carlo approximation of conditional expectations.
 \end{itemize}
\newpage

\section{Introduction}

  Let $Z$ be a $N(0,1)$ random variable. The distribution of
 $X:=|Z|$ is the so-called half-normal distribution. It will
be denoted $HN(0,1)$ and its density function is
$$f_X(x)=\sqrt{\frac{2}{\pi}}\exp\left\{-\frac12x^2\right\}I_{[0,+\infty[}(x).
$$

A general half-normal distribution $HN(\xi,\eta)$ is obtained from
$HN(0,1)$ by a location-scale transformation: $HN(\xi,\eta)$ is the
distribution of $Y=\xi+\eta X$.\par

The classical paper Daniel (1959) introduces half-normal plots
and the half-normal distribution, a special case of the folded and truncated normal distributions (see
Johnson, Kotz and Balakrishnan (1994)). Bland and Altman (1999) and  Bland (2005)
propose a so-called half-normal method to deal with relationships
between measurement error and magnitude, with applications in
medicine.
Pewsey (2002) uses the maximum likelihood principle to estimate the
parameters, and presents a brief survey on the general half-normal
distribution, its relations with other well-known distributions and
its usefulness in the analysis of highly skew data.  Pewsey (2004)
proposes bias-corrected versions of the maximum likelihood
estimators.
Nogales and Pérez (2011) deals with the problem of unbiased estimation for the general half-normal distribution. 

Here we consider the problem of equivariant estimation of the location and scale parameters, $\xi$
and $\eta$, but first we provide a brief review of results for
unbiased and maximum likelihood estimation appearing in the
literature.

The density function of $HN(\xi,\eta)$ is

$$f_Y(y)=\frac1{\eta} f_X\left(\frac{y-\xi}{\eta}\right)=\frac1{\eta} \sqrt{\frac2{\pi}}
\exp\left\{-\frac12 \left(\frac{y-\xi}{\eta}\right)^2
\right\}I_{[\xi,+\infty[}(y).
$$
It is readily shown that
$$E(Y)=\xi+\eta\sqrt{\frac{2}{\pi}}\qquad\text{and}\qquad \mbox{Var}(Y)=\frac{\pi-2}{\pi}\eta^2.
$$

Let us recall a lemma  from (Nogales and Pérez (2011)). We write $\Phi$ for the standard normal
cumulative distribution function.

\begin{mylem} \rm \label{l1}
Let $X_i=|Z_i|$, $1\le i\le n$, where $Z_1,\dots,Z_n$ is a sample of the standard normal distribution $N(0,1)$. Let $c_n:=E(X_{1:n})$, where $X_{1:n}$ denotes the minimun of $X_1,\dots,X_n$.

{\rm (i)} $c_n=\int_0^{\infty}(2-2\Phi(t))^n\,dt$.

{\rm (ii)} For $n\ge 1$, $c_n\le\frac1n\sqrt{\frac{\pi}{2}}\le
\Phi^{-1}\left(\frac12+\frac1{2n}\right)$.\end{mylem}

Let $Y_1,\dots,Y_n$ be a sample of size $n$ from a general half-normal distribution with unknown parameters, $\xi$ and $\eta$.
$Y_{1:n}$ denotes the minimum of $Y_1,\dots,Y_n$. From the factorization criterion, we obtain that $(\sum_{i=1}^nY_i^2,\sum_{i=1}^nY_i,Y_{1:n})$ is a sufficient statistic. Indeed, it is minimal sufficient, although not complete. With the notations of the lemma, we write $Y_i=\xi+\eta X_i$. Notice that $Y_{1:n}=\min_iY_i=\xi+\eta X_{1:n}$ and $E(Y_{1:n})=\xi+\eta c_n$.

The next proposition (Nogales and Pérez (2011)) yields unbiased
estimators of the location and scale parameters, $\xi$ and $\eta$.
Both estimators are $L$-statistics and functions of the cited
minimal sufficient statistic.

\begin{myprop}  \rm
Let $Y_1,\dots,Y_n$ be a sample of size $n$ from a general half-normal distribution with unknown parameters, $\xi$ and $\eta$.

{\rm (i)} $\widetilde\xi:=\frac{\sqrt{\frac{2}{\pi}}Y_{1:n}-
c_n{\bar Y}}{\sqrt{\frac{2}{\pi}}-c_n}$ is an unbiased estimator of
the location parameter~$\xi$.

{\rm (ii)} $\widetilde\eta:=\frac{{\bar
Y}-Y_{1:n}}{\sqrt{\frac{2}{\pi}}-c_n}$ is an unbiased estimator of
the scale parameter $\eta$ whose distribution does not depend on
$\xi$.

\end{myprop}

\begin{myrema}\rm  We also have that the sample mean ${\bar Y}$ is an unbiased estimator of the mean $\xi+\eta\sqrt{\frac{2}{\pi}}$. Moreover, an unbiased estimator of $\eta^2$ is
$$\frac{\pi}{\pi-2}\,S^2,
$$
where $S^2:=\frac1{n-1}\sum_{i=1}^n(Y_i-{\bar Y})^2$ is the sample
variance; notice that its distribution does not depend on $\xi$.
${\bar Y}$ and $S^2$ also are functions of the sufficient statistic
given above. The reader is referred to Nogales and Pérez (2011) for
these and other results about unbiased estimation of the parameters
of the general half-normal distribution. $\Box$ \end{myrema}

\begin{myrema}\rm  Pewsey (2002) provides maximum likelihood estimates for each of the parameters $\xi$ and $\eta$:
$$\widehat{\xi}:=Y_{1:n},\quad \widehat{\eta}:=\left(\frac1n\sum_{i=1}^n(Y_i-Y_{1:n})^2\right)^{1/2}
$$
A large sample based bias-correction is used in Pewsey (2004) to
improve the performance of the maximum likelihood estimators
$\widehat\xi$ and $\widehat\eta$. $\Box$ \end{myrema}

\section{A Monte Carlo method to approximate conditional expectations}
In this section, we describe a natural Monte Carlo method to compute conditional expectations based on a theorem of Besicovitch on differentiation of measures. It will be used in the next section to approximate the minimum risk equivariant (MRE) estimator of the location parameter $\xi$ because its expression involves two conditional expectations not easy to compute.
  
We first recall briefly a theorem of Besicovitch (1945, 1946) for differentiation of measures (see, for instance, Corollary 2.14 of Mattila (1995)). This theorem extend to Radon measures the classical Lebesgue Differentiation Theorem. 

\begin{mytheo}[Besicovitch (1945, 1946)] Let $\lambda$ be a Radon measure on $\mathbb R^n$, and $f:\mathbb
R^n\rightarrow\mathbb R$ a locally $\lambda$-integrable function. Then
$$\lim_{r\downarrow 0}\frac1{\lambda(B_r(x))}\int_{B_r(x)}f\,d\lambda=f(x)
$$
for $\lambda$-almost all $x\in\mathbb R^n$, where $B_r(x)$ denotes
the ball of center $x$ and radius $r>0$ for the norm
$\|\cdot\|_\infty$ on $\mathbb R^n$.
\end{mytheo}

Now let  $(\Omega,\mathcal A,P)$ be a probability space,
$X:(\Omega,\mathcal A,P)\rightarrow\mathbb R^n$ be an $n$-dimensional
random variable and $Y:(\Omega,\mathcal A,P)\rightarrow\mathbb R$ be a real random variable with finite mean. The conditional expectation $E(Y|X)$ is defined as a random variable on $\mathbb R^n$ such that
$\int_{X^{-1}(B)}Y\, dP=\int_BE(Y| X)dP^X$ for any Borel set $B$ in
$\mathbb R^n$, where $P^X$ denotes the probability distribution of
$X$.

Although the existence of the conditional expectation is
guaranteed via the Radon-Nikodym theo\-rem, its computation is,
generally, involved. Nevertheless, according to the previous result, for $P^X$-almost every
$x\in\mathbb R^n$,
$$\lim_{\epsilon\downarrow0}\frac1{P^X(B_\epsilon(x))}\int_{X^{-1}(B_\epsilon(x))}Y(\omega)\,dP(\omega)=
\lim_{\epsilon\downarrow0}\frac1{P^X(B_\epsilon(x))}\int_{B_\epsilon(x)}E(Y|X=x')\,dP^X(x')=E(Y|X=x)
$$

By the Strong Law of Large Numbers, for
almost every
sequence $(\omega_i)$ in $\Omega$, we have
\begin{gather*}
P^X(B_\epsilon(x))=\lim_k\frac1k\sum_{i=1}^kI_{B_\epsilon(x)}(X(\omega_i))\\ \mbox{ and }\\
\int_{B_\epsilon(x)}E(Y|X=x')\,dP^X(x')=\lim_k\frac1k\sum_{i=1}^kI_{B_\epsilon(x)}(X(\omega_i))Y(\omega_i)
\end{gather*}
where $I_A$ denotes the indicator function of $A$. Observe that, for every $\epsilon>0$, the rate of convergence is $1/\sqrt{n}$.

Hence, we have proved the following result:

\begin{mytheo} Let $(\Omega,\mathcal A,P)$ be a probability space,
$X:(\Omega,\mathcal A,P)\rightarrow\mathbb R^n$ be an $n$-dimensional
random variable and $Y:(\Omega,\mathcal A,P)\rightarrow\mathbb R$ be a
real random variable with finite mean. Then, for $P^X$-almost every
$x\in\mathbb R^n$ and almost every
sequence $(\omega_i)$ in $\Omega$, we have
$$E(Y|X=x)=\lim_{\epsilon\downarrow0}\lim_k\frac{\sum_{i=1}^kI_{B_\epsilon(x)}(X(\omega_i))Y(\omega_i)}{\sum_{i=1}^kI_{B_\epsilon(x)}(X(\omega_i))}."
$$
\end{mytheo}

This theorem yields a means of approximating the conditional
expectation of $Y$ given $X$. The following simple example
illustrates the method. 

\begin{myexa}\rm 
Let $(X,Y)$ be a bivariate normal random variable with null mean such that $\Var(X)=\Var(Y)=1$ and $\Cov(X,Y)=0.5$.
In this case, there is no need for an approximation to the
conditional expectation of $Y$ given $X=x$ because it is $x/2$. The
conditional distribution of $Y$ given $X=x$ is $N(\frac12
x,\frac12\sqrt{3})$. Applying the proposed method to evaluate
$E(Y|X=1)$, given a small $\epsilon>0$, we may choose a sample
$(x_i,y_i)_{1\le i\le k}$ from the joint distribution of $X$ and $Y$
and approximate $E(Y|X=1)$ by
\begin{gather*}\frac{\sum_{i=1}^kI_{[1-\epsilon,1+\epsilon]}(x_i)
y_i}{\sum_{i=1}^kI_{[1-\epsilon,1+\epsilon]}(x_i)}.\tag{1}
\end{gather*}
Taking $\epsilon=0.1, 0.01$ and samples from the joint
distribution of $X$ and $Y$ with sample sizes $k$ large enough to
obtain $m=m(k)=\sum_{i=1}^kI_{[1-\epsilon,1+\epsilon]}(x_i)=100, 1000, 5000$, we obtained the approximations for
$E(Y|X=1)$ summarized in Table 1 and Figure 1; 100 replications of each simulation have been conducted to obtain the table and the figure. Namely, taking $m=1000$, for instance, the value 0.493947 appearing in the table as an approximation of $E(Y|X=1)$ when $\epsilon=0.1$ is the mean of the 100 values of the quotient (1) obtained after 100 replications of the experiment of choosing a $k$-sized sample $(x_i,y_i)_{1\le i\le k}$ of the joint distribution of $(X,Y)$, $k$ being large enough to get $m=m(k)=1000$. Table 1 also includes the ``mean squared error'' (MSE) calculated from these 100 values: the format used for a typical entry in the table is $E(Y|X=1)\pm \mbox{MSE}$. The box-plot of the figure describes the distribution of these 100 values (a dotted red line represents the mean). 
\begin{center}\begin{tabular}{|c|c|c|c|c|c|}\hline
$m$ & 100 & 1000 & 5000 \\\hline 
$\epsilon=0.1$ & $0.505885 \ \&\ 0.006395$ & $0.493947 \ \&\ 0.000815$ & $0.497892 \ \&\ 0.000128$ \\\hline
$\epsilon=0.01$ & $0.503655 \ \&\ 0.007165$ & $0.499826 \ \&\ 0.000716$ &
$0.499471 \ \&\ 0.000150$ \\\hline
\end{tabular}\vspace{2ex}\\
{\small Table 1. Approximation  of $E(Y|X=1) \ \&\  \mbox{MSE}$ as a function of the number of simulations, $m$, for $\epsilon=0.1, 0.01$.}\vspace{2ex}\\
\scalebox{.8}{\includegraphics[width=200pt,height=200pt]{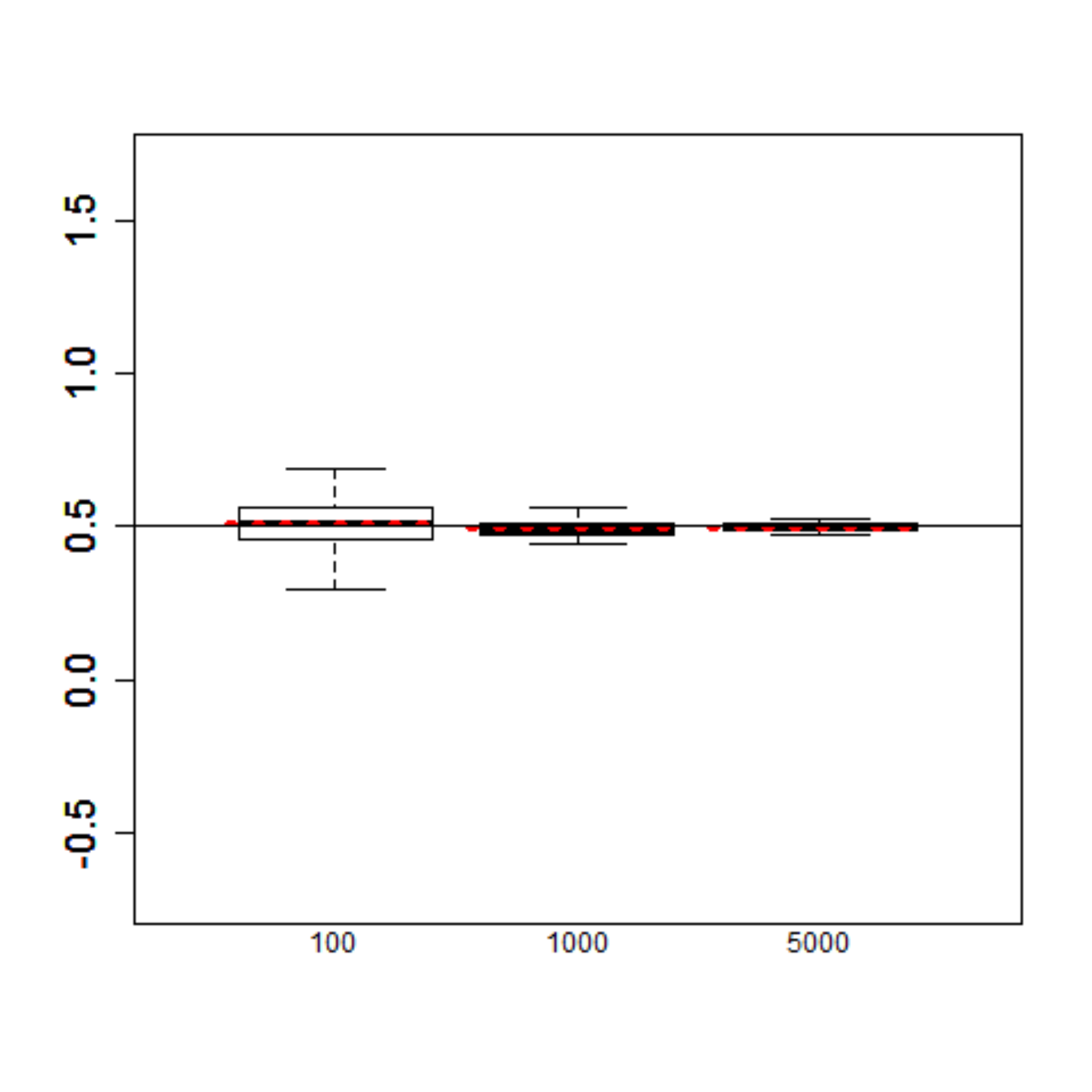}}\scalebox{.8}{\includegraphics[width=200pt,height=200pt]{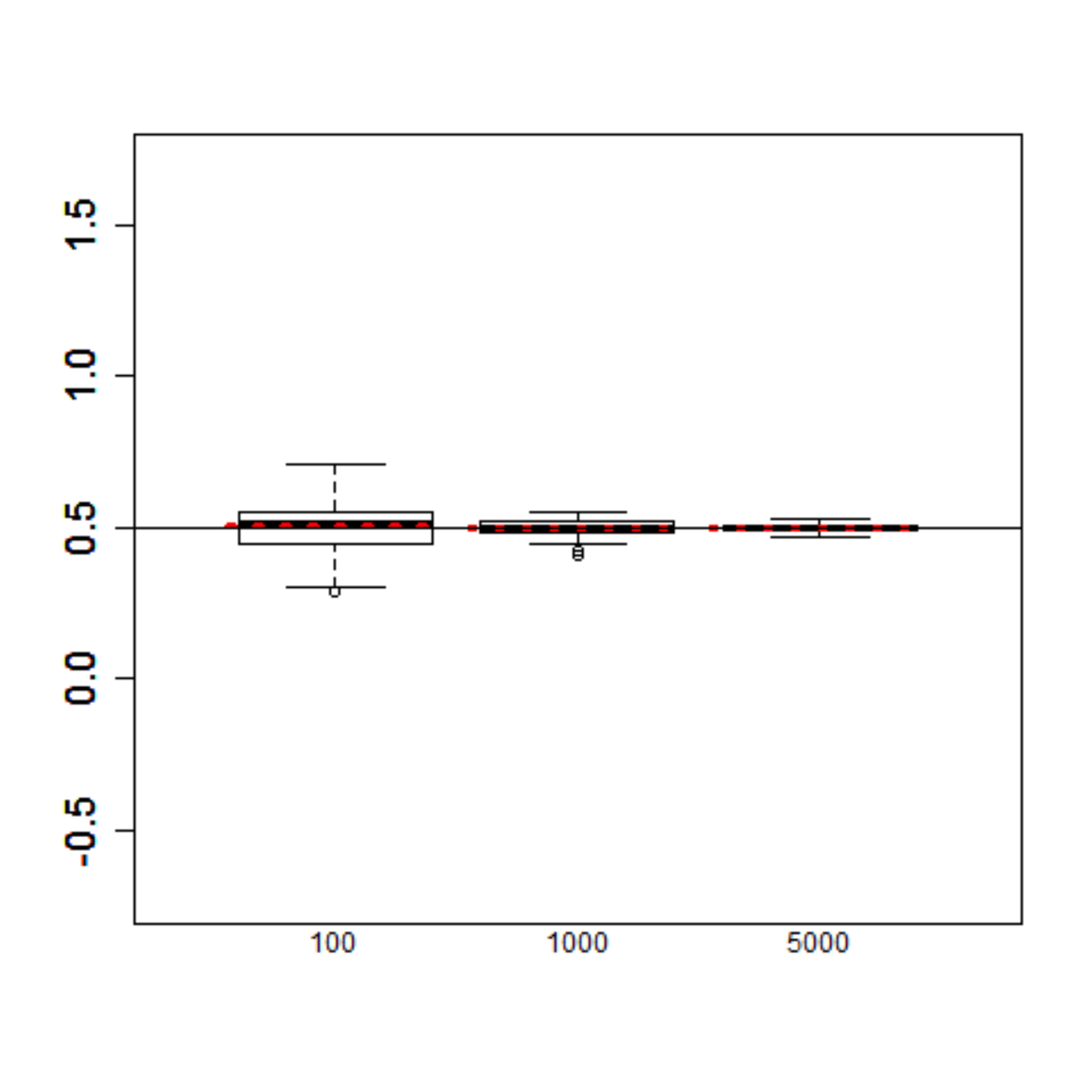}}\vspace{-3ex}\\
{\small Figure 1. Box plots of the approximations  of $E(Y|X=1)$}\vspace{-0.9ex}\\ {\small as a
function of the number of simulations, $m$, for
$\epsilon=0.1$ and $\epsilon=0.01$.}\end{center}
\end{myexa}

\begin{myrema}\rm 
The described method of Monte Carlo approximation to the conditional expectation $E(Y|X=x)$ is based on the naive idea that one can approximate it from a sample $(x_i,y_i)_{1\le i\le n}$ by the mean of the $y_i$ corresponding to points $x_i$ lying in a narrow neighborhood of $x$. From a probabilistic point of view, the method has been justified by the mentioned theorem of Besicovitch on differentiation of measures.
When the joint density of $X$ and $Y$ is known, $E(Y| X=x)$ is the mean of the conditional distribution of $Y$ given $X=x$, and the problem of compute a conditional expectation is reduced to the problem of computing a mean. Notice that the existence of a joint density is not required by the method and it could be specially useful when densities are not available or are not easy to compute (see the next example). $\Box$
\end{myrema}
 
 \begin{myexa}\rm \  (Example 1, continuation)
 A similar simulation study has been performed to appro\-xi\-mate the conditional expectation $E(V|U=0.5)$, where $V=\sin(X\cdot Y)$ and $U=\cos(X^2+Y^2)$; the obtained results are:
 
 \begin{center}\begin{tabular}{|c|c|c|c|c|c|}\hline
 $m$ & 100 & 1000 & 5000 \\\hline 
 $\epsilon=0.1$ & $0.127650 \ \&\ 0.001890$ & $0.127280 \ \&\ 0.000202$ & $0.124169 \ \&\ 0.000025$ \\\hline
 $\epsilon=0.01$ & $0.123063 \ \&\ 0.001620$ & $0.125869 \ \&\ 0.000153$ &
 $0.1252856 \ \&\ 0.000031$ \\\hline
 \end{tabular}\vspace{2ex}\\
 {\small Table 2. Approximation  of $E(V|U=0.5) \ \&\  S^2$ ($S^2$ is the sample variance) as a\\ function of the number of simulations, $m$, for $\epsilon=0.1, 0.01$.}\vspace{2ex}\\
 \scalebox{.8}{\includegraphics[width=200pt,height=200pt]{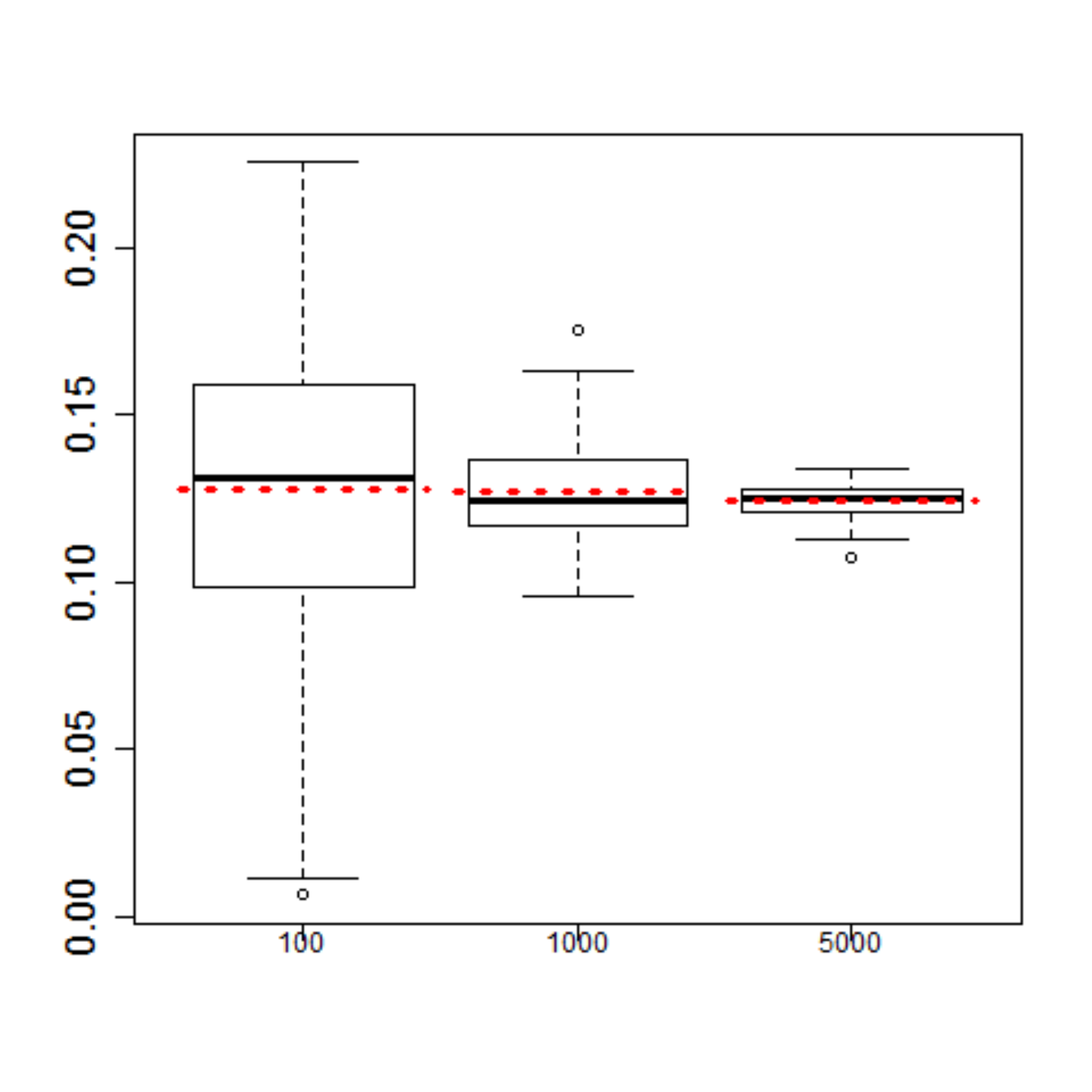}}\scalebox{.8}{\includegraphics[width=200pt,height=200pt]{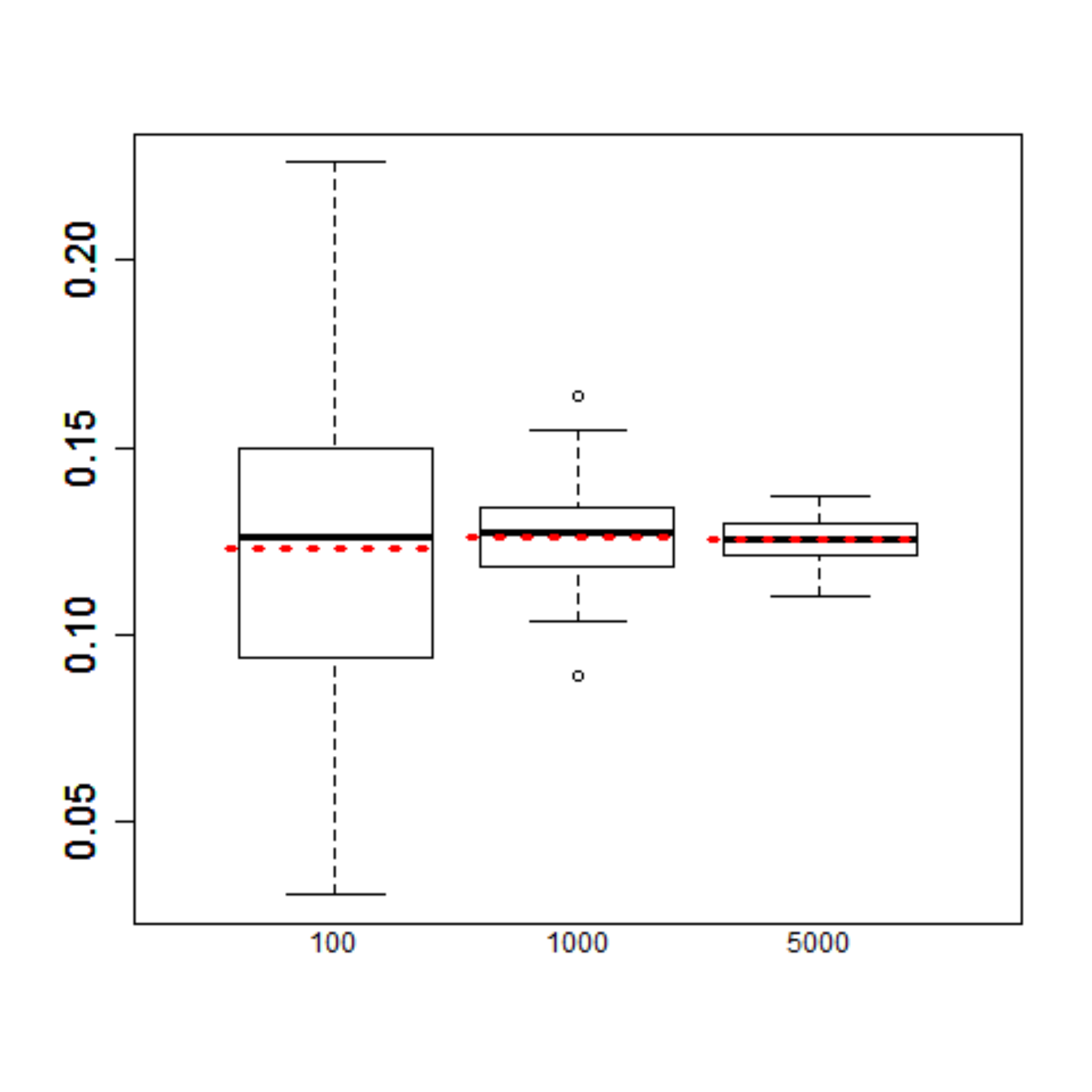}}\vspace{-3ex}\\
 {\small Figure 2. Box plots of the approximations  of $E(V|U=0.5)$}\vspace{-0.9ex}\\ {\small as a
 function of the number of simulations, $m$, for
 $\epsilon=0.1$ and $\epsilon=0.01$.}\end{center}
 \end{myexa}

       \begin{myrema}\rm
In a classical statistical framework, we can provide additional guarantees on the method, since the obtained Monte Carlo approximation to the conditional expectation $E(Y|X=x)$ coincides with the value at the point $x$ of the kernel estimator (the Nadaraya-Watson estimator) of the regression curve $y=E(Y|X=x)$ for the kernel $K(x)=I_{[-1,1]}(x)$ (see Nadaraya (1989), p. 115). From this point of view, $\epsilon$ plays the role of the bandwidth parameter. We  refer to Härdle (1992, Ch. 5) for a detailed discussion on the important problem of the choice of the bandwidth.  
$\Box$
\end{myrema}

       \begin{myrema}\rm
As it is pointed out to us by the referees, in a Bayesian setting a similar idea has been in use in recent years to generate an approximate sample from the posterior distribution given $x$ assuming that the likelihood function is easy to sample. This proceeds by sampling values $\theta_i$  from the prior distribution and $x_i$ from the distribution of the data given $\theta_i$, and accepting those parameters $\theta_i$ such that $x_i$ is in the ball $B_\epsilon(x)$ centered at $x$ of radius $\epsilon>0$. In fact, from a sample of size $k$ we can approximate the posterior probability given $x$ of a subset $T$ of the parameter space by
$$\frac{\sum_{i=1}^kI_{B_\epsilon(x)}(x_i)I_T(\theta_i)}{\sum_{i=1}^kI_{B_\epsilon(x)}(x_i)}.
$$
We also can approximate the posterior mean given $x$ of a function $f$ of the parameter by
$$\phantom{XXXX}\frac{\sum_{i=1}^kI_{B_\epsilon(x)}(x_i)f(\theta_i)}{\sum_{i=1}^kI_{B_\epsilon(x)}(x_i)}.
\qquad\Box$$

\end{myrema}

       \begin{myrema}\rm
In this paper, the main application of the Monte Carlo method for the approximation of conditional expectations is given in the next section to approximate the estimation of the location parameter of the general half-normal distribution, because it is defined in terms of a quotient of two not-easily-computable parameter-free conditional expectations given a $(n-1)$-dimensional statistic $U$.  Some ``curse of dimensionality problem" appears when $n$ is large because, in this case, it is not easy to find large samples of points lying in a small ball centered at a point $U(y)$. This is why we had to modify the Monte Carlo method for the approximation of conditional expectations taking advantage of the underlying distribution of $Y$ (the general half-normal distribution) and the invariance properties of $U$. This could become an important scholium of the paper, as the ideas used here could be useful to deal with the ``curse of dimensionality problem"  in similar situations.    
$\Box$
\end{myrema}

\section{Equivariant estimation of the location parameter of the ge\-ne\-ral half-normal distribution}

In this section we consider the problem of determining the minimum
risk equivariant estimator of the location parameter $\xi$ of the
general half-normal distribution $HN(\xi,\eta)$ when the scale
parameter $\eta$ is unknown. We cannot provide an explicit
expression for this estimator, since it is described in terms of two
conditional expectations that had to be approximated by simulation. 

To achieve this goal, an R program was developed based on the method of computing conditional expectations described in the previous section. In fact, the method has been slightly modified to solve a sort of ``curse of dimensionality" problem.

We consider the scale-location family of densities
$$f_{(\xi,\eta)}(y_1,...,y_n)=\frac{1}{\eta^n}f\left(\frac{y_1-\xi}{\eta},...,
\frac{y_n-\xi}{\eta}\right),$$ where
$$f(y_1,...,y_n)=
\left(\frac{2}{\pi}\right)^{\frac{n}{2}}
\exp\left\{-\frac{1}{2}\sum_{i=1}^n y_i^2\right\}
I_{[0,+\infty[}(y_{1:n}).$$ This family remains invariant under
transformations of the form
$g_{a,b}(y_1,...,y_n)=(a+by_1,...,a+by_n)$, $a\in\mathbb{R}$, $b>0$.

To estimate the location parameter $\xi$ when the scale parameter
$\eta$ is unknown, we have the next result, a direct consequence of
classical equivariant estimation theory (see Lehmann (1983)). First, recall that an estimator $T$ of the location parameter is equivariant if $T(a+bx_1,\dots,a+bx_n)=a+bT(x_1,\dots,x_n)$, for all $a\in\mathbb R$ and all $b>0$.

\begin{myprop} 
When the loss function $W_2(x;\xi,\eta)=\eta^{-2}(x-\xi)^2$ is considered,
the MRE estimator $\mathring{\xi}$ of $\xi$ is
$$\mathring{\xi}=T^*_0-(\rho\circ U) T^*_1
$$
where
\begin{gather*}
T^*_0=\bar Y,\quad T^*_1=\frac1n\sum_{i=1}^n|Y_i-\bar Y|\\
U=\left(\frac{Y_1-Y_n}{Y_{n-1}-Y_n},\dots,\frac{Y_{n-2}-Y_n}{Y_{n-1}-Y_n},\frac{Y_{n-1}-Y_n}{|Y_{n-1}-Y_n|}\right),\\
\rho=\frac{E_{\xi=0,\eta=1}(T^*_0
T^*_1|U)}{E_{\xi=0,\eta=1}({T^*_1}^2|U)}
\end{gather*}
\end{myprop}

\begin{myrema}\rm
$T^*_0$ can be replaced by any other equivariant estimator of $\xi$ (i.e., satisfying $T^*_0(a+by_1,\dots,a+by_1)=a+b T^*_0(y_1,\dots,y_1)$ for
every $a\in\mathbb R$, $b>0$),
and $T^*_1$ can be replaced by any positive estimator of $\eta$
satisfying $T^*_1(a+by_1,\dots,a+by_1)=b T^*_1(y_1,\dots,y_1)$ for
every $a\in\mathbb R$, $b>0$. $\Box$ \end{myrema}

A simulation study has been performed to investigate the behavior of the
minimum risk equivariant estimator $\mathring{\xi}$. In it, we
used 100 simulations with sample sizes $n=100,1000,5000$ from the
$HN(10,4)$ distribution, obtaining the results summarized in Table 3
and Figure 3 (see below how we have made use of the method of approximation of conditional expectations to obtain the values of the Tables 3 and 4).

\begin{center}\begin{tabular}{|c|c|c|c|c|c|}\hline
$m$ & 100 & 1000 & 5000 \\\hline 
$\epsilon=0.1$ & $9.412753 \ \&\ 2.307873$ & $9.517691 \ \&\ 1.881063$ & $9.732626 \ \&\ 0.158635$ \\\hline
$\epsilon=0.01$ & $9.603969 \ \&\ 0.687243$ & $9.274164 \ \&\ 3.501158$ &
$9.867600 \ \&\ 0.027826$ \\\hline
\end{tabular}\vspace{2ex}\\
{\small Table 3. Approximations  of $\mathring{\xi} \ \&\  \mbox{MSE}$ as a function of the number of simulations, $m$, for $\epsilon=0.1, 0.01$.}\vspace{2ex}\\
\scalebox{.8}{\includegraphics[width=200pt,height=200pt]{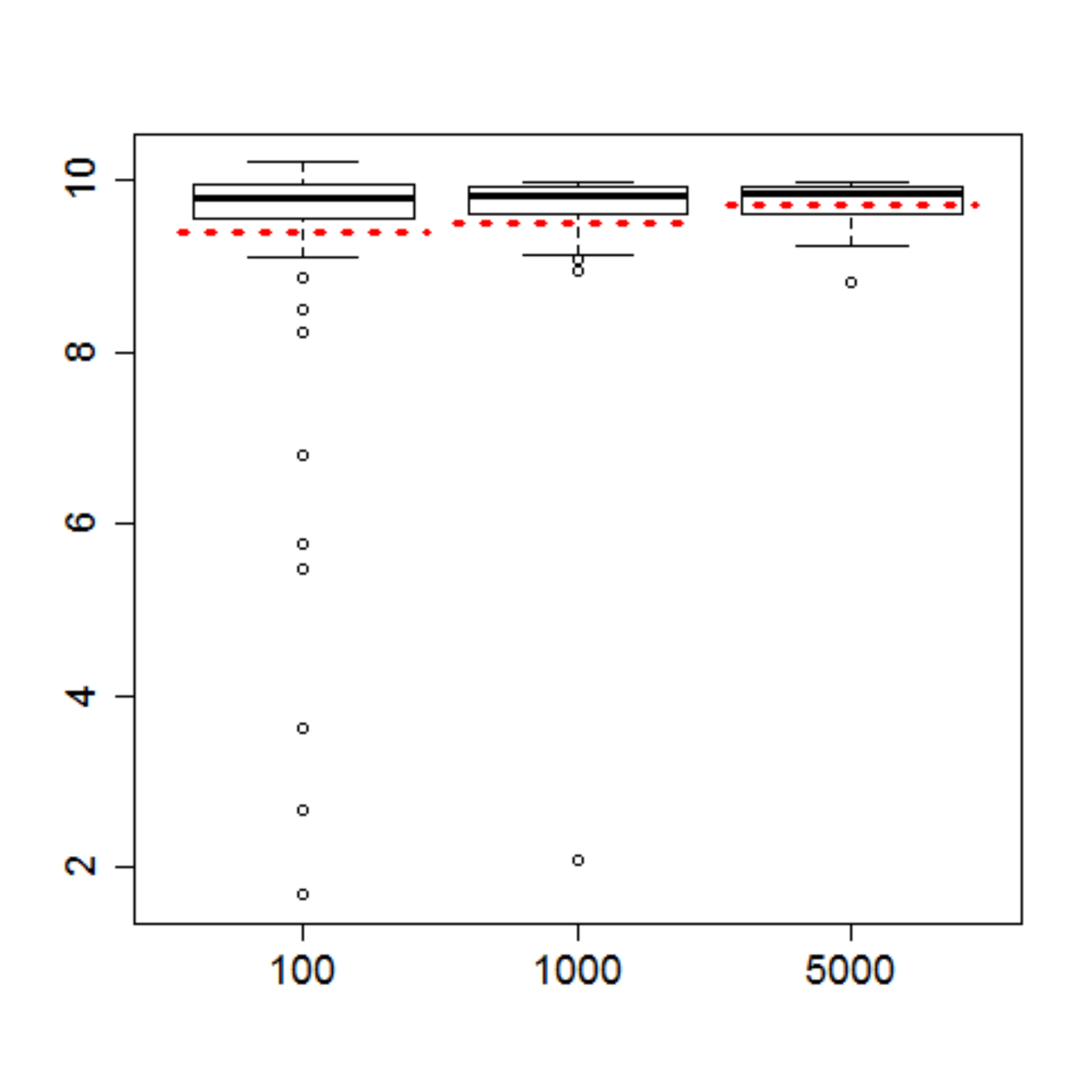}}\scalebox{.8}{\includegraphics[width=200pt,height=200pt]{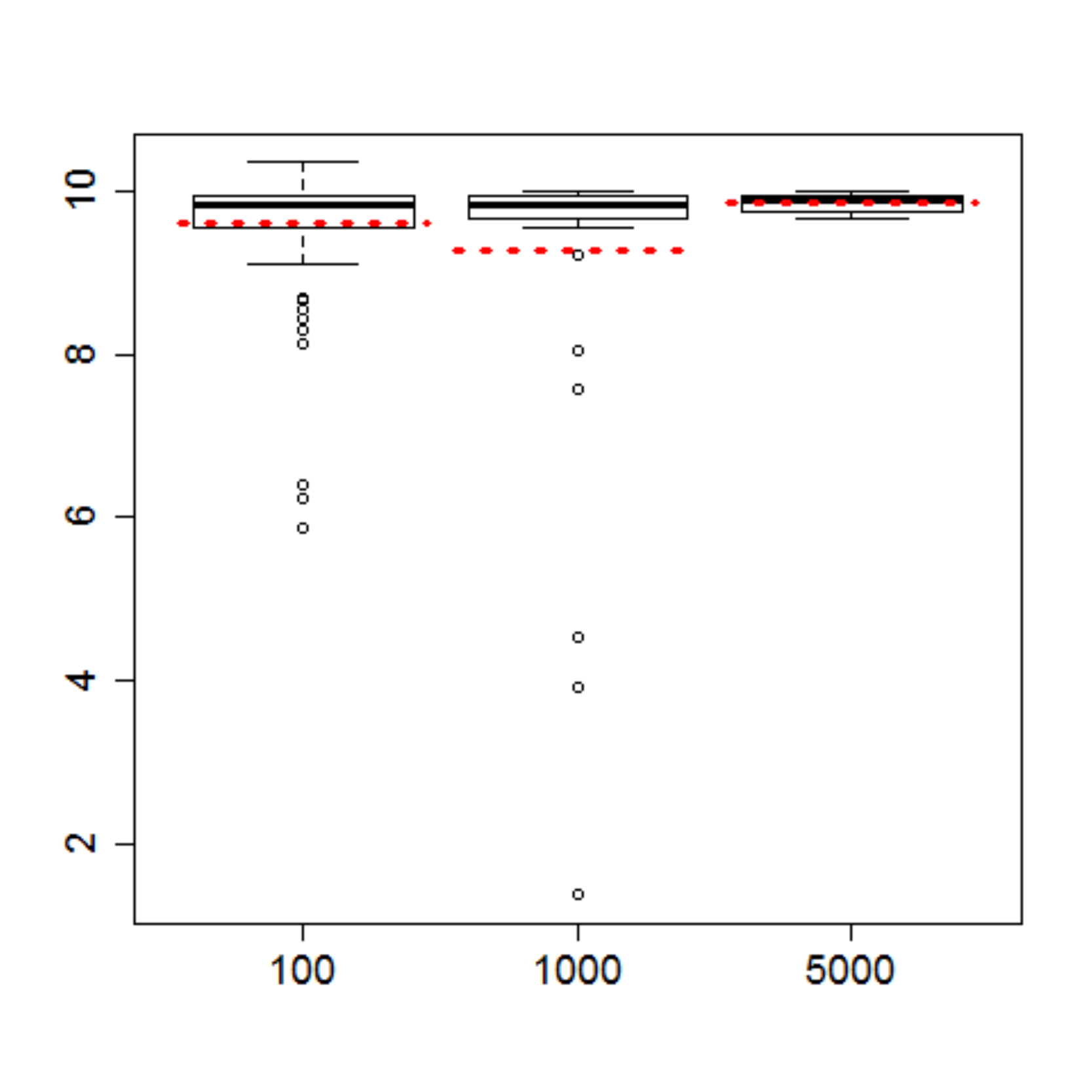}}\vspace{-2ex}\\
{\small Figure 3. Box plots of the approximations  of $\mathring{\xi} \ \&\ \mbox{MSE}$}\vspace{-0.9ex}\\ {\small as a
function of the number of simulations, $m$, for
$\epsilon=0.1$ and $\epsilon=0.01$.}\end{center}

To compare the behavior of the unbiased estimator $\tilde{\xi}$, the
maximum likelihood estimator $\hat{\xi}$ and the minimum risk
equivariant estimator $\mathring{\xi}$, we used 100
simulations with sample sizes $n=100,1000,5000$ from the
$HN(10,4)$ distribution, obtaining the results summarized in Table 4
and Figure 4:

\begin{center}
\begin{tabular}{|c|c|c|c|c|c|c|}\hline
& $m$ & 100 & 1000 & 5000 \\\hline 
\multirow{2}{*}{$\tilde{\xi}$}& $\epsilon=0.1$ & $9.997350 \ \&\ 0.003289$ & $9.999356 \ \&\ 0.000020$ & $10.000823 \ \&\ 0.000001$ \\\cline{2-5}
& $\epsilon=0.01$ & $9.999662 \ \&\ 0.003443$ & $9.999989 \ \&\ 0.000025$ &
$10.001050 \ \&\ 0.000002$ \\\hline
\multirow{2}{*}{$\hat{\xi}$}& $\epsilon=0.1$ & $10.047256 \ \&\ 0.005455$ & $10.004383 \ \&\ 0.000039$ & $10.000823 \ \&\ 0.000001$ \\\cline{2-5}
& $\epsilon=0.01$ & $10.049128 \ \&\ 0.005752$ & $10.005005 \ \&\ 0.000050$ &
$10.001050 \ \&\ 0.000002$ \\\hline
\multirow{2}{*}{$\mathring{\xi}$}& $\epsilon=0.1$ & $9.412753 \ \&\ 2.307873$ & $9.517691 \ \&\ 1.881063$ & $9.732626 \ \&\ 0.158635$ \\\cline{2-5}
& $\epsilon=0.01$ & $9.603969 \ \&\ 0.687243$ & $9.274164 \ \&\ 3.501158$ &
$9.867600 \ \&\ 0.027826$ \\\hline
\end{tabular}\vspace{2ex}\\
{\small Table 4. Approximations  of $\tilde{\xi} \ \&\  \mbox{MSE}$, $\hat{\xi} \ \&\  \mbox{MSE}$ and $\mathring{\xi} \ \&\  \mbox{MSE}$ as a function\\ of the number of simulations, $m$, for $\epsilon=0.1, 0.01$.}\vspace{2ex}
\end{center}

\begin{figure}\begin{center}
{\boldmath{$m=100$}}\\ \scalebox{.8}{\includegraphics[width=200pt,height=200pt]{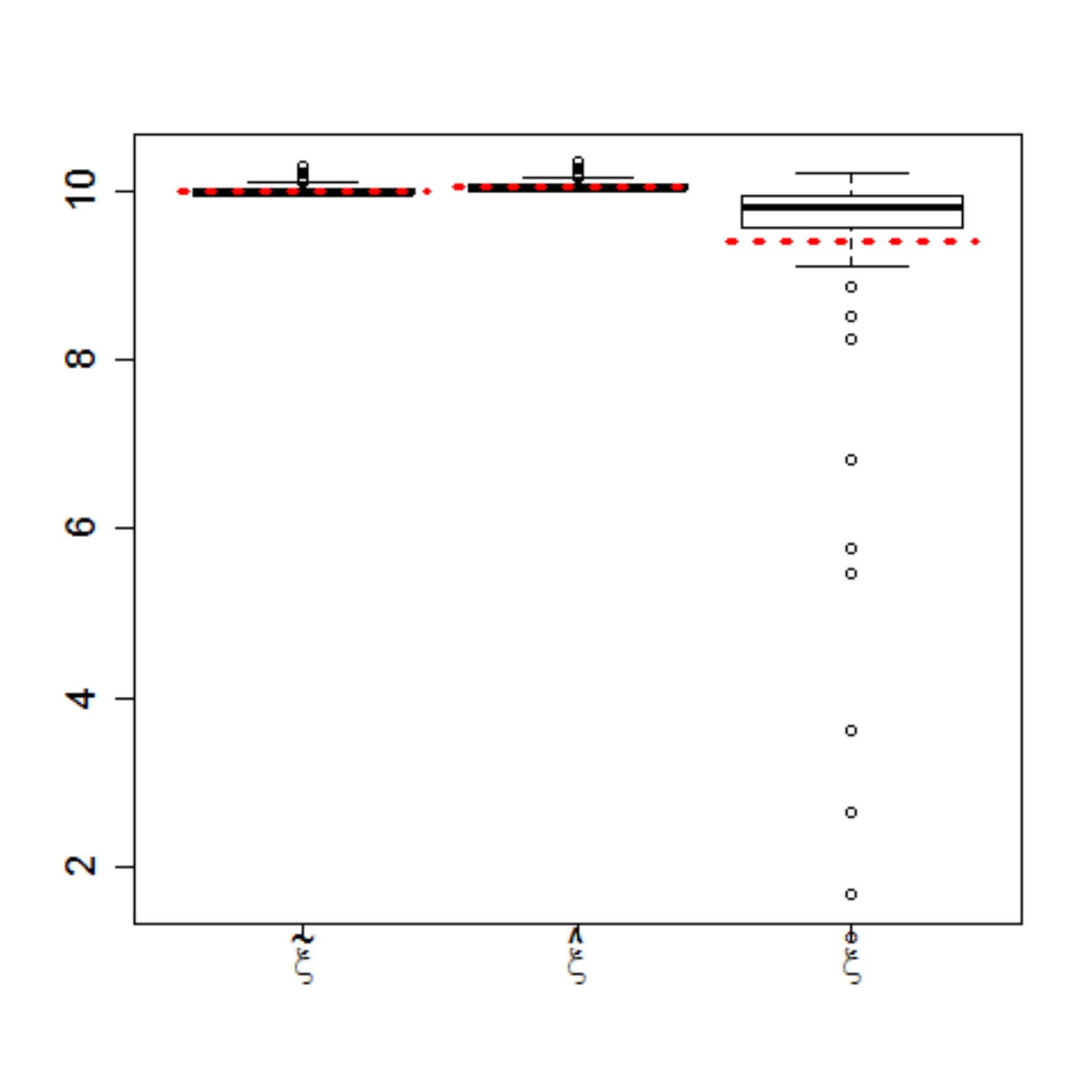}}\scalebox{.8}{\includegraphics[width=200pt,height=200pt]{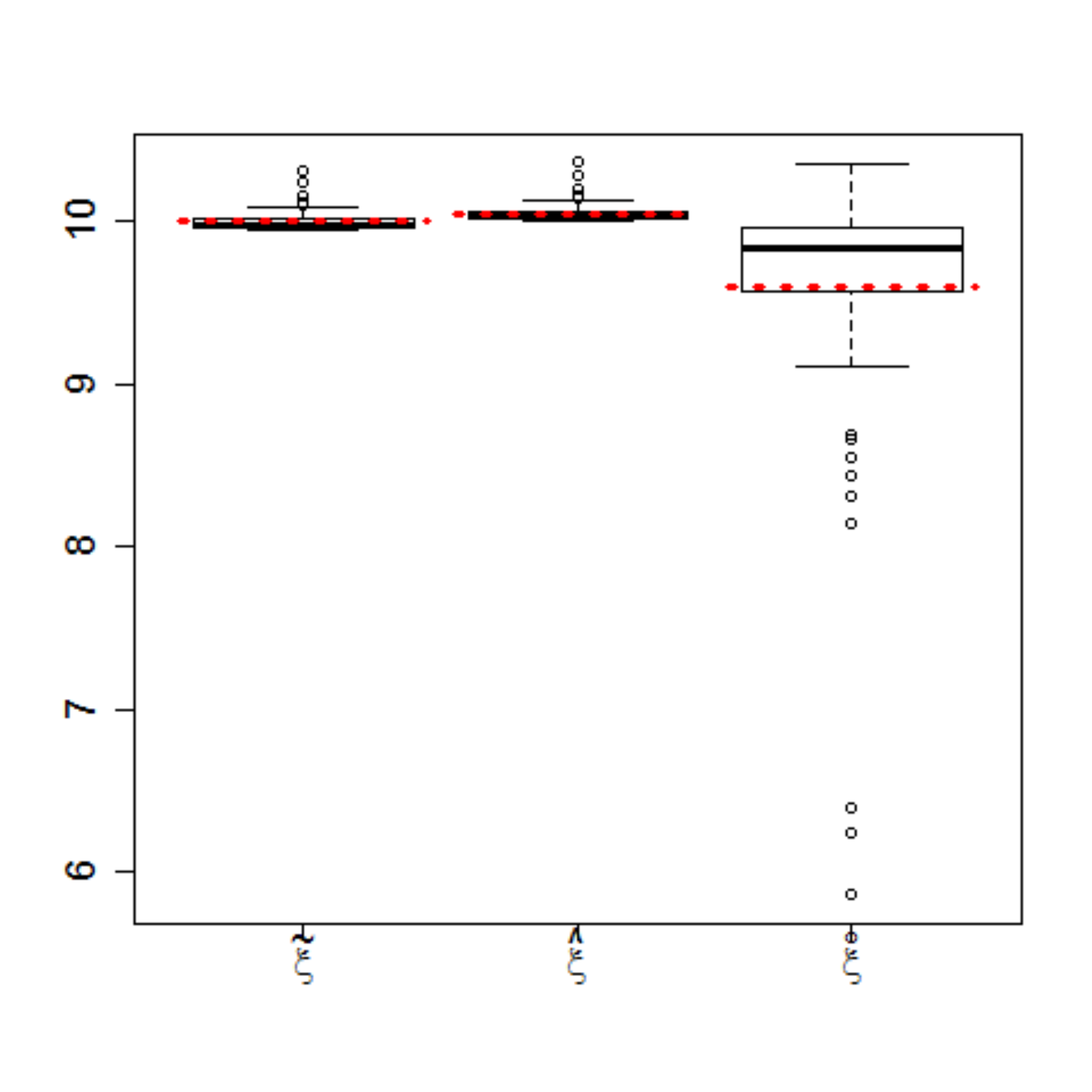}}\vspace{1ex}\\
{\boldmath{$m=1000$}}\\
\scalebox{.8}{\includegraphics[width=200pt,height=200pt]{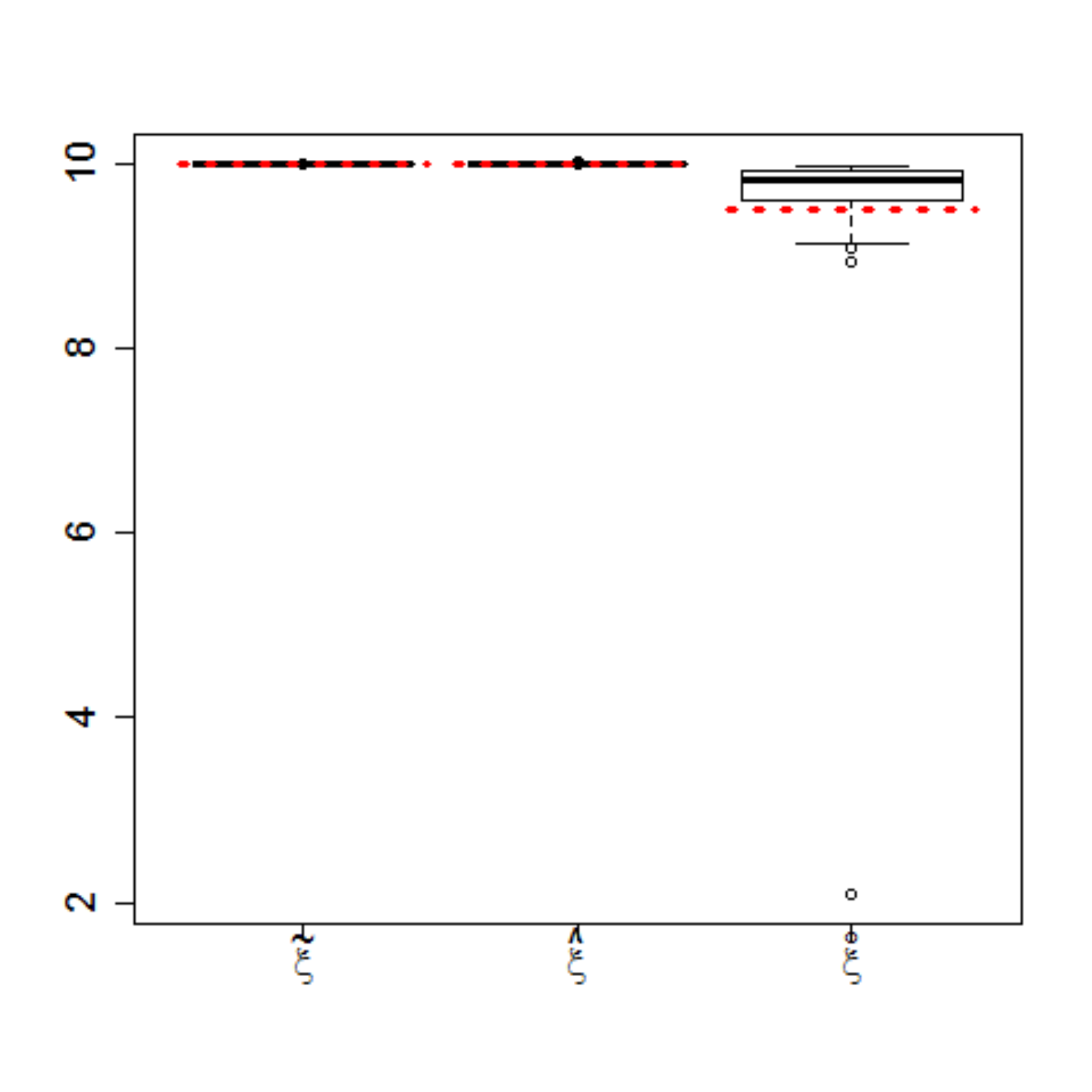}}\scalebox{.8}{\includegraphics[width=200pt,height=200pt]{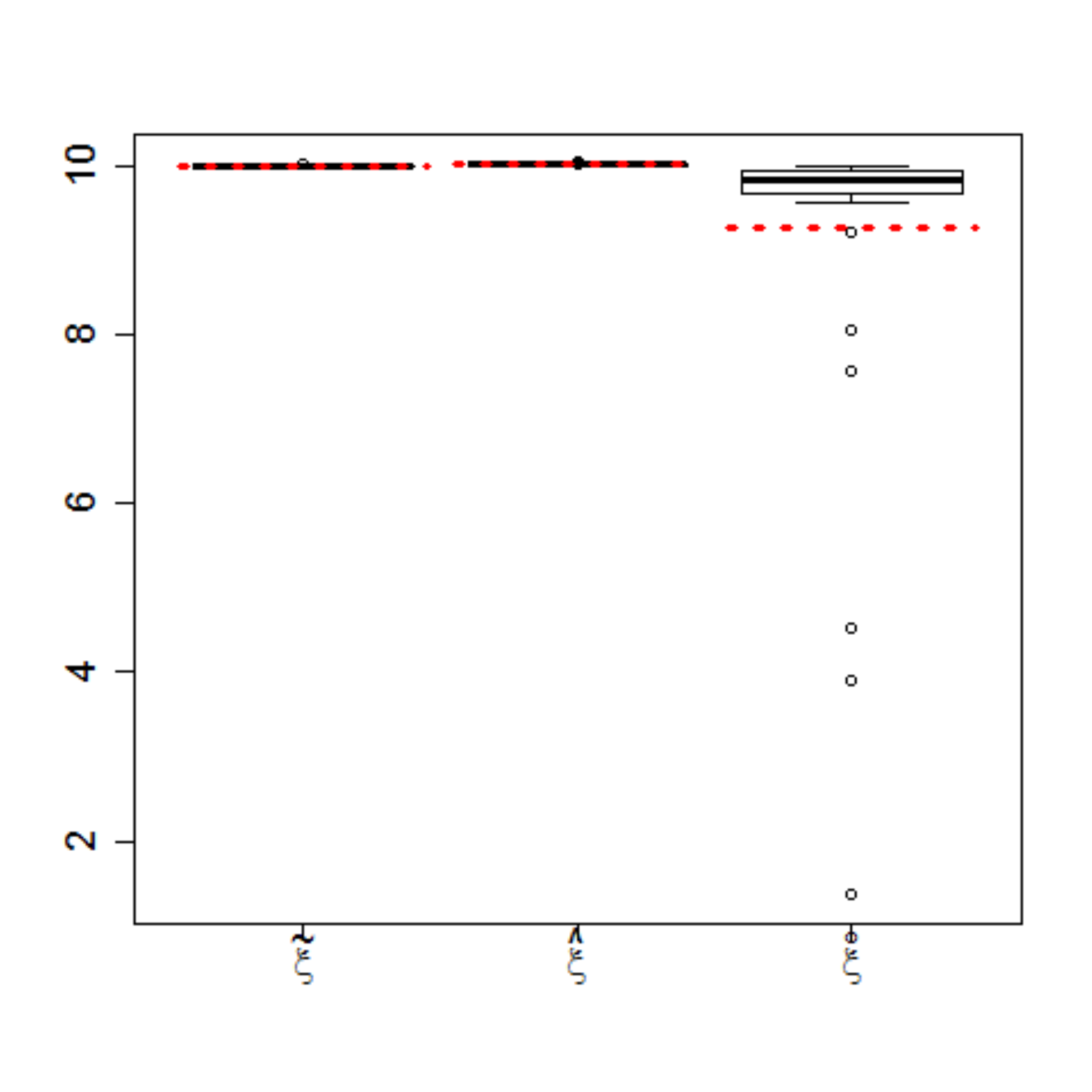}}\vspace{1ex}\\
{\boldmath{$m=5000$}}\\
\scalebox{.8}{\includegraphics[width=200pt,height=200pt]{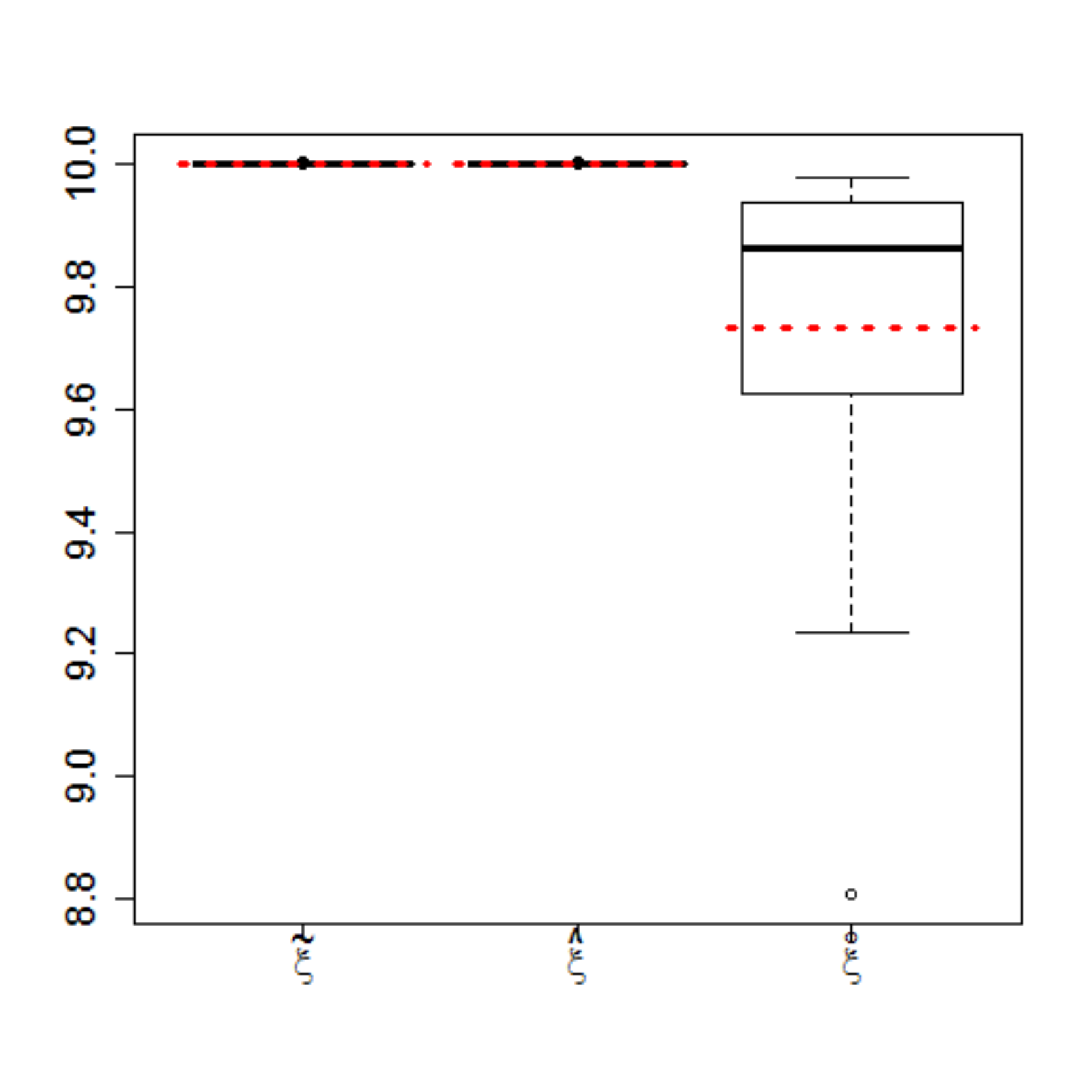}}\scalebox{.8}{\includegraphics[width=200pt,height=200pt]{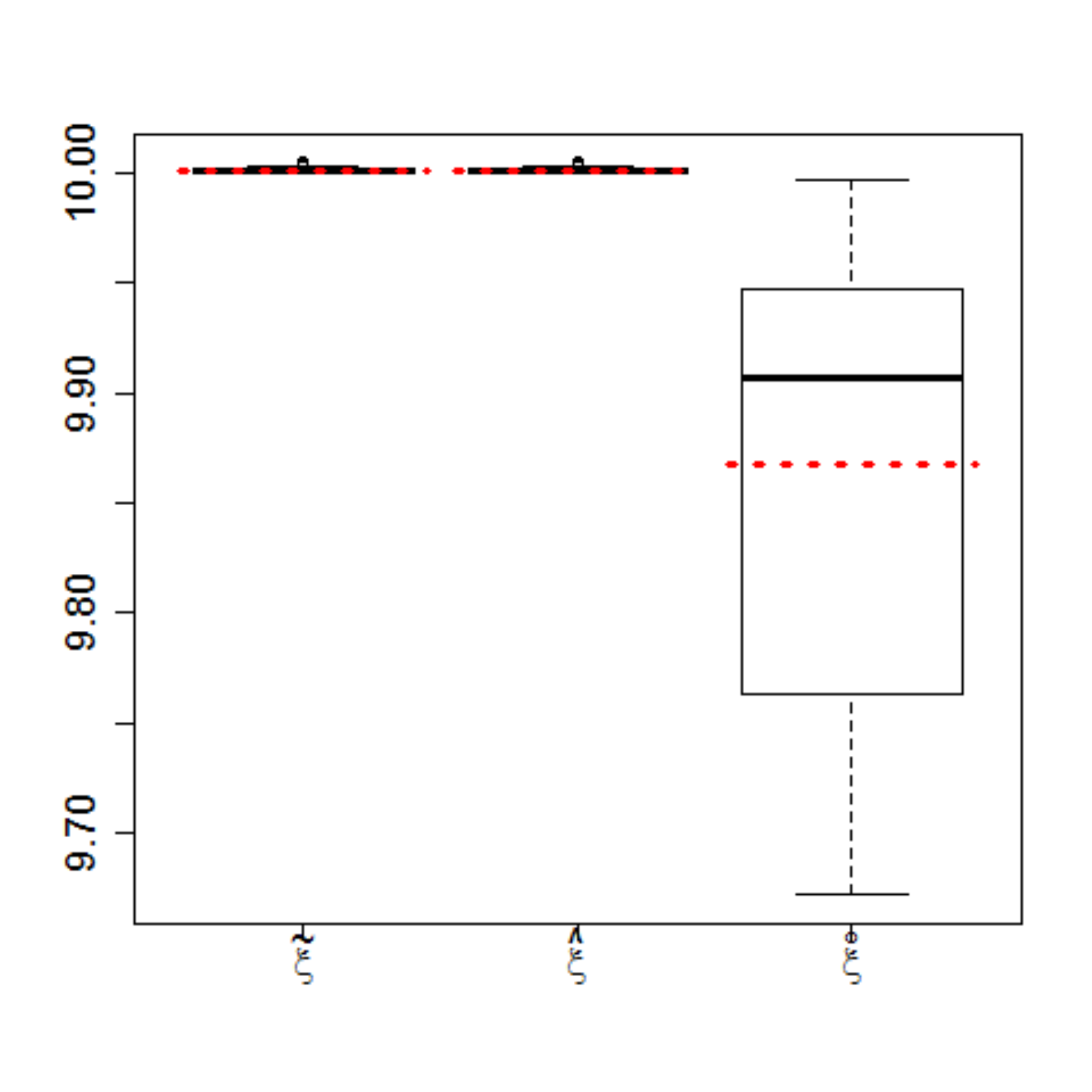}}\vspace{-2ex}\\
{\small Figure 4. Box plots of the approximations  of  $\tilde{\xi}$, $\hat{\xi}$ and $\mathring{\xi}$ as a
function of the} \vspace{-0.9ex}\\ {\small number of simulations, $m=100,1000,5000$, for
$\epsilon=0.1$ and $\epsilon=0.01$.}\end{center}\end{figure}

Table 4 and Figure 4 illustrate the biased character of the maximum
likelihood estimator $\hat{\xi}$ and the minimum risk equivariant
estimator $\mathring{\xi}$.

Let us describe in more details the ideas used in these simulations. For a sample
$y=(y_1,\dots,y_n)$, $n=100, 1000, 5000$, of the distribution
$HN(10,4)$, we have
$$\rho(U(y))=\lim_{\epsilon\to 0}\frac{N_\epsilon}{D_\epsilon}
$$
where
\begin{gather*}
N_\epsilon=\int_{A_\epsilon(y)}f(y')dy',\quad D_\epsilon=\int_{A_\epsilon(y)}g(y')dy',\\
f(y')=T^*_0(y') T^*_1(y') \exp\left\{-\frac12\|y'\|_2^2\right\},\quad g(y')=T^*_1(y')^2 \exp\left\{-\frac12\|y'\|_2^2\right\},\\
A_\epsilon(y)=\{y'\in[0,10]^n\colon \max_{1\le i\le
n-1}|U_i(y')-U_i(y)|\le\epsilon\}.
\end{gather*}
Now, take a sample $S$ of $A_\epsilon(y)$ and approximate $N_\epsilon$
and $D_\epsilon$ by 
$$\frac1{\mbox{card\,}(S)}\sum_{y'\in S}f(y')\quad\mbox{and}\quad \frac1{\mbox{card\,}(S)}\sum_{y'\in S}g(y'),$$
respectively. So,
$\rho(U(y))$ can be approximated by
$$C(y):=\frac{\sum_{y'\in S}f(y')}{\sum_{y'\in S}g(y')}
$$
and $\mathring{\xi}(y)$ is approximated by $D(y):=T^*_0(y)-C(y) T^*_1(y)$.

To approximate $C(y)$, a first idea would be to divide the interval
$[0,10]$ in multiple subintervals of small length $\epsilon>0$ and
consider the grid in the interval $[0,10]^n$ formed by the $n$-power
set of the ends of these subintervals (we have restricted ourselves
to the interval [0,10] because the
functions $ f (y) $ and $ g (y) $ are almost null when one of the coordinates of the
vector $y$ is greater than 10). The sample $S$ would then be formed by the
grid nodes that are in $A_\epsilon$. The main problem with this
approach is that the size $m$ of the sample $S$ is very small: it
becomes smaller as $n$ increases, because of the so-called ``curse of dimensionality" problem. In order to avoid this problem and obtain a sample size $m$ large enough for $S$ (given $n$, we take $m=100 n$), we have used the following algorithm, a modification of the described Monte Carlo method to approximate conditional expectations that hinges on the use of the invariance
of $U$ under scale and location transformations. \label{11} Namely:
 
\begin{itemize}
\item[Step A.] Let $n\in\mathbb N$ and be $y=(y_1,\dots,y_n)$ a $n$-sized sample of the distribution $HN(10,4)$. For $1\le i\le n-2$, let $a_i:=\frac{y_{i}-y_{n}}{y_{n-1}-y_{n}}$ and take $0<\epsilon<\min\{0.1, \min_{1\le i\le n-2}|a_i|\}$. 
\begin{itemize}
\item[Step A.1.] At this stage we choose coordinatewise at random $100\cdot n$ vectors $v^{(j)}=(v^{(j)}_1,\dots,v^{(j)}_n)$, $1\le j\le 100 n$, in $\mathbb R^n$ such that $\max_{1\le i\le n-1}|U_i(v^{(j)})-U_i(y)|\le \epsilon$ as follows:

\begin{itemize}
\item[A.1.1.] Make $j=1$. 
\item[A.1.2.] Take $v^{(j)}_{n-1},v^{(j)}_{n}$ at random in $[0,10]$ such that $v^{(j)}_{n-1}-v^{(j)}_{n}$ has the same sign as $y_{n-1}-y_{n}$. (So, the last coordinates of $U(v^{(j)})$ and $U(y)$ are the same). 

\item[A.1.3.] For $1\le i\le n-2$ take $v^{(j)}_i$ at random on the interval determined by $v^{(j)}_n+(v^{(j)}_{n-1}-v^{(j)}_{n})(a_i-\epsilon)$ and
$v^{(j)}_n+(v^{(j)}_{n-1}-v^{(j)}_{n})(a_i+\epsilon)$. (So $|U_i(v^{(j)})-U_i(y)|\le \epsilon$).

\item[A.1.4.] Make $j=j+1$ a go back to Step A.1 until $100 n$ vectors $v^{(j)}=(v^{(j)}_1,\dots,v^{(j)}_n)$, $1\le j\le 100 n$ are obtained.
     \end{itemize}

\item[Step A.2.] Since the vectors $v^{(j)}=(v^{(j)}_1,\dots,v^{(j)}_n)$, $1\le j\le 100 n$, do not lie necessarily in $[0,10]^n$ (so neither in $A_\epsilon(y)$), we can make some random location-scale transformations to put them into $[0,10]^n$. These transformations do not modify the required fact that  $\max_{1\le i\le n-1}|U_i(v^{(j)})-U_i(y)|\le \epsilon$.
\begin{itemize}
\item[A.2.1.] If $v^{(j_0)}_{i_0}<0$ for some $i_0,j_0$, we define $u^{(j)}_i=v+v^{(j)}_i$,  $1\le i\le n$,
$1\le j\le 100 n$, where $v$ is choosen at random between 
$-\min_{1\le i\le n, 1\le j\le 100 n}v^{(j)}_i$ and $1-\min_{1\le i\le n, 1\le j\le 100 n}v^{(j)}_i$. Otherwise,  $u^{(j)}_i=v^{(j)}_i$,  $1\le i\le n$, $1\le j\le 100 n$. 

\item[A.2.2.] Each vector $u^{(j)}$ is divided by $\max_{1\le i\le n}u^{(j)}_i$ and multiplied by a random number choosen in $[0,10]$ to obtain the vector $w^{(j)}$.

\item[A.2.3.] Take $S=\{w^{(j)}\colon 1\le j\le 100 n\}$ and approximate $C(y)$ by
$$\frac{\sum_{j=1}^{100n}f(w^{(j)})}{\sum_{j=1}^{100n}g(w^{(j)})}
$$
and $D(y)$ by $T^*_0(y)-C(y) T^*_1(y)$.
\end{itemize}
\end{itemize}
\item[Step B.] Finally,  following the process designed in Step A, we choose $k:=100$ random samples $y^{(i)}$ of size $n$ from the $HN(10,4)$ distribution and approximate the mean and the mean squared error of $\mathring{\xi}$ by
$$\frac1k\sum_{i=1}^kD(y^{(i)})\quad\text{and}\quad \frac1k\sum_{i=1}^k(D(y^{(i)})-10)^2,
$$
respectively, and we construct a box-plot with the values $D(y^{(i)})$.
\end{itemize}

\begin{myrema}\rm
Notice that both $\hat{\xi}$ and $\tilde{\xi}$ are equivariant estimators of the location parameter $\xi$. So they have greater risk for the loss function $W_2$ than $\mathring{\xi}$. Hence, in the previous simulation study, the MSE of $\mathring{\xi}$ should have been smaller than the MSE of $\hat{\xi}$ and $\tilde{\xi}$. That has not been the case because, for the MRE estimator, we have not real estimates of $\xi$, but approximations of these estimates obtained by a modification of the Monte Carlo method of computing the conditional expectations appearing as the numerator and denominator of a quotient. But this is a possible issue to approximate minimum risk estimations of a location parameter, and a possible way to avoid the ``curse of dimensionality problem''. $\Box$
       \end{myrema}
       
\begin{myrema}\rm
Although less interesting from the perspective of real applications, for completeness we now consider
the problem of estimating the scale parameter $\xi$ when the
location parameter $\eta$ is known, say $\eta=\eta_0$. In this case, 
 the joint density of $Y_1,\dots,Y_n$ is
$$f_\xi(y_1,\dots,y_n)=\frac1{\eta_0^n} \sqrt{\frac2{\pi}}^{\;n}
\exp\left\{-\frac1{2\eta_0^2} \sum_{i=1}^n(y_i-\xi)^2
\right\}I_{[\xi,+\infty[}(y_{1:n}),
$$
where $y_{1:n}:=\min\{y_1,\dots,y_n\}$. This family remains
invariant under translations of the form
$g_a(y_1,\dots,y_n)=(y_1-a,\dots,y_n-a)$.

The equivariant estimator of minimum mean squared error of the
location parameter $\xi$ is
$$T_1={\bar Y}-\frac{\eta_0}{\sqrt{2\pi n}}\frac{\exp\left\{-\frac{n}{2\eta_0^2}\left(Y_{1:n}-{\bar Y}\right)^2\right\}}
{\Phi\left[\frac{\sqrt{n}}{\eta_0}\left(Y_{1:n}-{\bar
Y}\right)\right]}.
$$

In fact, for the loss function $W'_2(\xi,x)=(x-\xi)^2$, the MRE
estimator of the location parameter $\xi$ is the Pitman estimator
$$T_1(y_1,\dots,y_n)=\frac{\int_{-\infty}^{+\infty}uf_0(y_1-u,...,y_n-u)du}
{\int_{-\infty}^{+\infty}f_0(y_1-u,...,y_n-u)du}.
$$
For $y\in\mathbb {R}^n$, we write $\bar y$ for the mean of
$y_1,\dots, y_n$. After some algebraic manipulations, we obtain:

\begin{gather*}
\int_{-\infty}^{+\infty}uf_0(y_1-u,...,y_n-u)du=\\
\begin{split}\left(\frac{\sqrt{2}} {\eta_0\sqrt{\pi}}\right)^n
&\exp\left\{-\frac{1}{2\eta_0^2}\left(\sum_{i=1}^n y_i^2-n{\bar
y}^2\right)\right\}\frac{\eta_0}{\sqrt{n}}\\
&\times\left[-\frac{\eta_0}{\sqrt{n}}\exp\left\{-\frac{n}{2\eta_0^2}(y_{1:n}
-{\bar y})^2\right\}+{\bar y}\sqrt{2\pi}\,
\Phi\left(\frac{\sqrt{n}}{\eta_0} (y_{1:n}-{\bar y})\right)\right]
\end{split}\end{gather*}
and
\begin{gather*}
\int_{-\infty}^{+\infty}f_0(y_1-u,...,y_n-u)du=\\
\left(\frac{\sqrt{2}}{\eta_0\sqrt{\pi}}\right)^n
\exp\left\{-\frac{1}{2\eta_0^2}\left(\sum_{i=1}^n y_i^2-n{\bar
y}^2\right)\right\}\frac{\eta_0}{\sqrt{n}} \sqrt{2\pi}\,
\Phi\left[\frac{\sqrt{n}}{\eta_0} (y_{1:n}-{\bar y})\right]
\end{gather*}
and the statement follows easily from these expressions. $\Box$
\end{myrema}

\section{Equivariant estimation of the scale parameter of the general half-normal distribution}

Unlike what happens with the location parameter $\xi$, for the scale
parameter $\eta$ an explicit expression for the MRE estimator is
obtained.

Recall that an estimator $T$ of the scale parameter $\eta$ is equivariant if $T(a+bx_1,\dots,a+bx_n)=bT(x_1,\dots,x_n)$, for all $a\in\mathbb R$ and all $b>0$.

\begin{myprop} 
When using the loss function $W_1(x;\xi,\eta)=\eta^{-2}(x-\eta)^2$,
the MRE estimator $\mathring{\eta}$ of $\eta$ is
$$\mathring{\eta}=\sqrt{\frac{n-1}{2}}\frac{\Gamma\left(\frac{n+1}{2}\right)}{\Gamma\left(\frac{n+2}{2}\right)}
\frac{t_{n+1}\left(\left[\sqrt{\frac{n(n+1)}{n-1}}\frac{\bar
Y-Y_{1:n}}{S},\infty\right[\right)}{t_{n+2}\left(\left[\sqrt{\frac{n(n+2)}{n-1}}\frac{\bar
Y-Y_{1:n}}{S},\infty\right[\right)} S.
$$
where $t_{n}$ denotes the Student's $t$-distribution with $n$
degrees of freedom, $S^2$ is the sample variance and $\Gamma$ denotes Euler's gamma function.
\end{myprop}

\begin{myproo} \rm
The MRE estimator of the scale parameter $\eta$, when using the
loss function $W_1$, is
$$\mathring{\eta}(y)=\frac{\int_0^{+\infty}v^nf'(vy'_1,...,vy'_{n-1})dv}
{\int_0^{+\infty}v^{n+1}f'(vy'_1,...,vy'_{n-1})dv},$$ where $f'$ is
the joint density when $\eta=1$ of $Y'_i:=Y_i-Y_n$, $1\le i\le n-1$,
and $y'_i:=y_i-y_n$, $1\le i\le n-1$.

Notice that

\begin{gather*}
f'(y'_1,...,y'_{n-1})=\int_{-\infty}^{+\infty}f(y_1+t,...,y_n+t)dt\\
=\left(\frac{2}{\pi}\right)^{\frac{n}{2}}\exp\left\{-\frac12\sum_{i=1}^ny_i^2+\frac{n}{2}{\bar y}^2\right\}\int_{-y_{1:n}}^\infty\exp\left\{-\frac{n}{2}(t+\bar y)^2\right\}dt\\
=\frac1{\sqrt{n}}\left(\frac{2}{\pi}\right)^{\frac{n}{2}}\exp\left\{-\frac12(n-1)S^2(y)\right\}\int_{\sqrt{n}(\bar y-y_{1:n})}^\infty\exp\left\{ -\frac12u^2\right\}du.
\end{gather*}

Hence, for $k\in\mathbb{N}$, applying Fubini's Theorem after a suitable change of variables in the inner integral,
\begin{gather*}
I_k(y):=\int_0^{\infty}v^kf'(vy'_1,...,vy'_{n-1})dv\\=\frac1{\sqrt{n}}\left(\frac{2}{\pi}\right)^{\frac{n}{2}}
\int_0^\infty v^k\exp\left\{-\frac12(n-1)v^2S^2(y)\right\}\int_{\sqrt{n}(\bar y-y_{1:n})}^\infty\exp\left\{ -\frac12u^2\right\}dudv\\
=\frac1{\sqrt{n}}\left(\frac{2}{\pi}\right)^{\frac{n}{2}}
\int_{\sqrt{n}(\bar y-y_{1:n})}^\infty J_k(t,y)dt.\end{gather*}
where
\begin{gather*}
J_k(t,y):=\int_0^\infty v^{k+1}\exp\left\{-\frac12v^2(t^2+(n-1)S^2(y))\right\} dv=\frac{2^{k/2}\Gamma\left(\frac{k+2}{2}\right)}{(t^2+(n-1)S^2(y))^{\frac{k+2}{2}}}.
\end{gather*}
where, for $t\ge \sqrt{n}(\bar y-y_{1:n})$, we have made the change of variables $w=\frac12v^2(t^2+(n-1)S^2(y))$.

So,
\begin{gather*}
I_k(y)=\frac1{\sqrt{n}}\left(\frac{2}{\pi}\right)^{\frac{n}{2}}2^{k/2}\Gamma\left(\frac{k+2}{2}\right)
\int_{\sqrt{n}(\bar y-y_{1:n})}^{\infty}\frac{dt}{(t^2+(n-1)S^2(y))^{\frac{k+2}{2}}}\\
=\frac{2^{\frac{n+k}{2}}\Gamma\left(\frac{k+1}{2}\right)}{\sqrt{n}\pi^{\frac{n-1}{2}}(n-1)^{\frac{k+1}{2}}S(y)^{k+1}} t_{k+1}\left(\left[\sqrt{\frac{n(k+1)}{n-1}}\frac{\bar y-y_{1:n}}{S(y)},\infty\right[\right).
\end{gather*}
 Finally
$$\mathring{\eta}(y)=\frac{I_n(y)}{I_{n+1}(y)}=\sqrt{\frac{n-1}{2}}\frac{\Gamma\left(\frac{n+1}{2}\right)}{\Gamma\left(\frac{n+2}{2}\right)}
\frac{t_{n+1}\left(\left[\sqrt{\frac{n(n+1)}{n-1}}\frac{\bar
y-y_{1:n}}{S(y)},\infty\right[\right)}{t_{n+2}\left(\left[\sqrt{\frac{n(n+2)}{n-1}}\frac{\bar
y-y_{1:n}}{S(y)},\infty\right[\right)} S(y).
$$
$\Box$ \end{myproo}

\begin{myrema}\rm  A simulation study has been performed to compare the behavior of the unbiased estimator
$\tilde\eta$, the maximum likelihood estimator $\hat\eta$ and the
MRE estimator $\mathring{\eta}$ using 1000 simulated random samples of size $n=10, 20, 30$ from the $HN(10,4)$
distribution. The results obtained for the means and the mean squared errors of the three estimators are presented in Table 5 and Figure 5 (as before, a dotted red line represents the mean).

\begin{center}\begin{tabular}{|c|c|c|c|}\hline
 $n$ & $\tilde{\eta}$  & $\hat{\eta}$  & $\mathring{\eta}$ \\\hline
 10 & $3.996009  \ \&\ 1.052443$  &  $3.520680 \ \&\ 0.952987$  & $3.568520 \ \&\ 0.929288$ \\\hline
 20 & $3.996575 \ \&\ 0.526328$ &  $3.760888 \ \&\ 0.458780$  & $3.795590 \ \&\ 0.450882$ \\\hline
 30 & $4.015727 \ \&\ 0.324161$ & $3.845478 \ \&\ 0.294937$ & $3.871677 \ \&\ 0.291209$ \\\hline
\end{tabular}\vspace{2ex}\\
{\small Table 5. Sample mean and MSE of the estimators calculated using 1000 random\\ samples
of size $n$ from the $HN(10, 4)$ distribution.}
\\
\scalebox{.8}{\includegraphics[width=200pt,height=200pt]{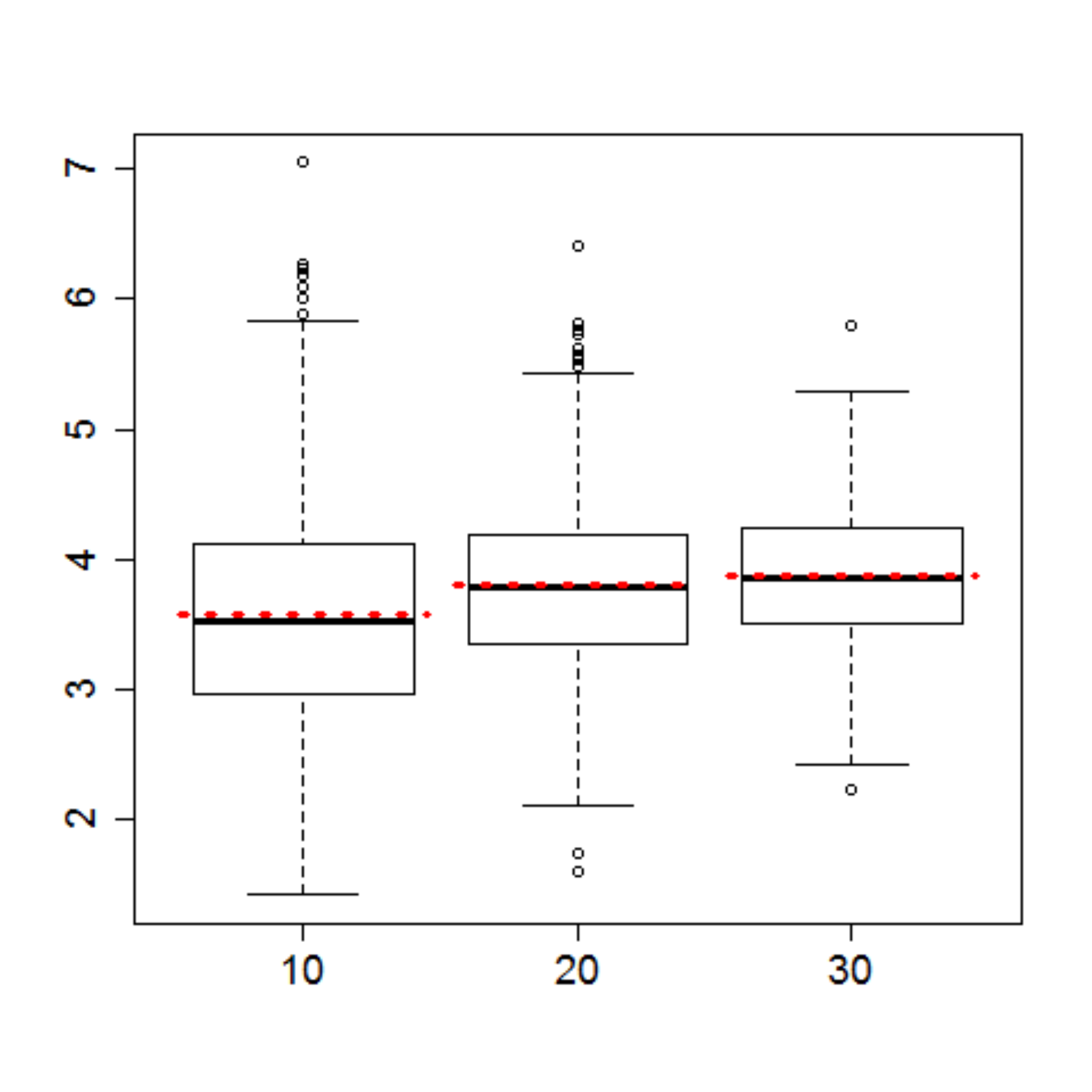}}\vspace{-2ex}\\
\scalebox{.8}{\includegraphics[width=150pt,height=200pt]{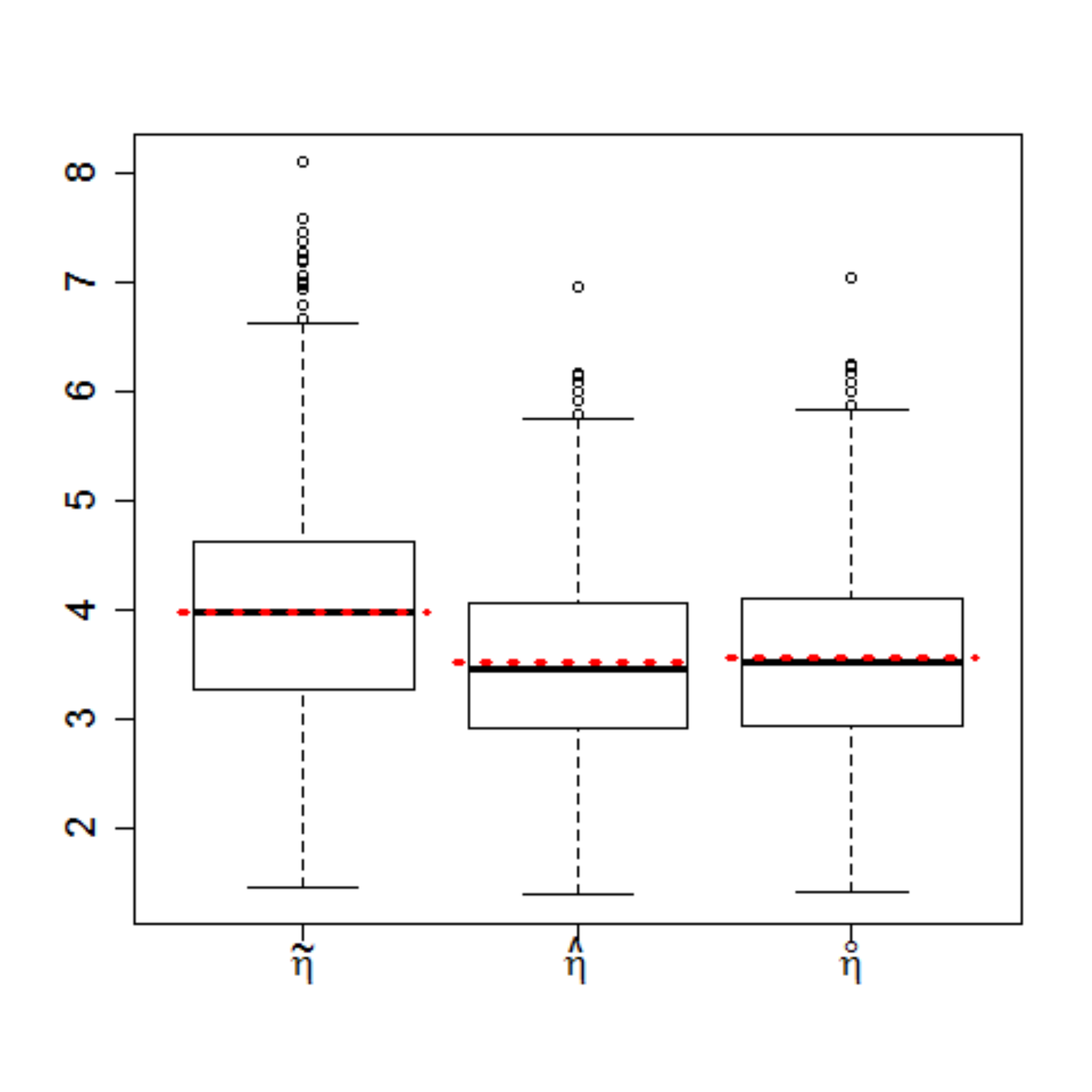}}
\scalebox{.8}{\includegraphics[width=150pt,height=200pt]{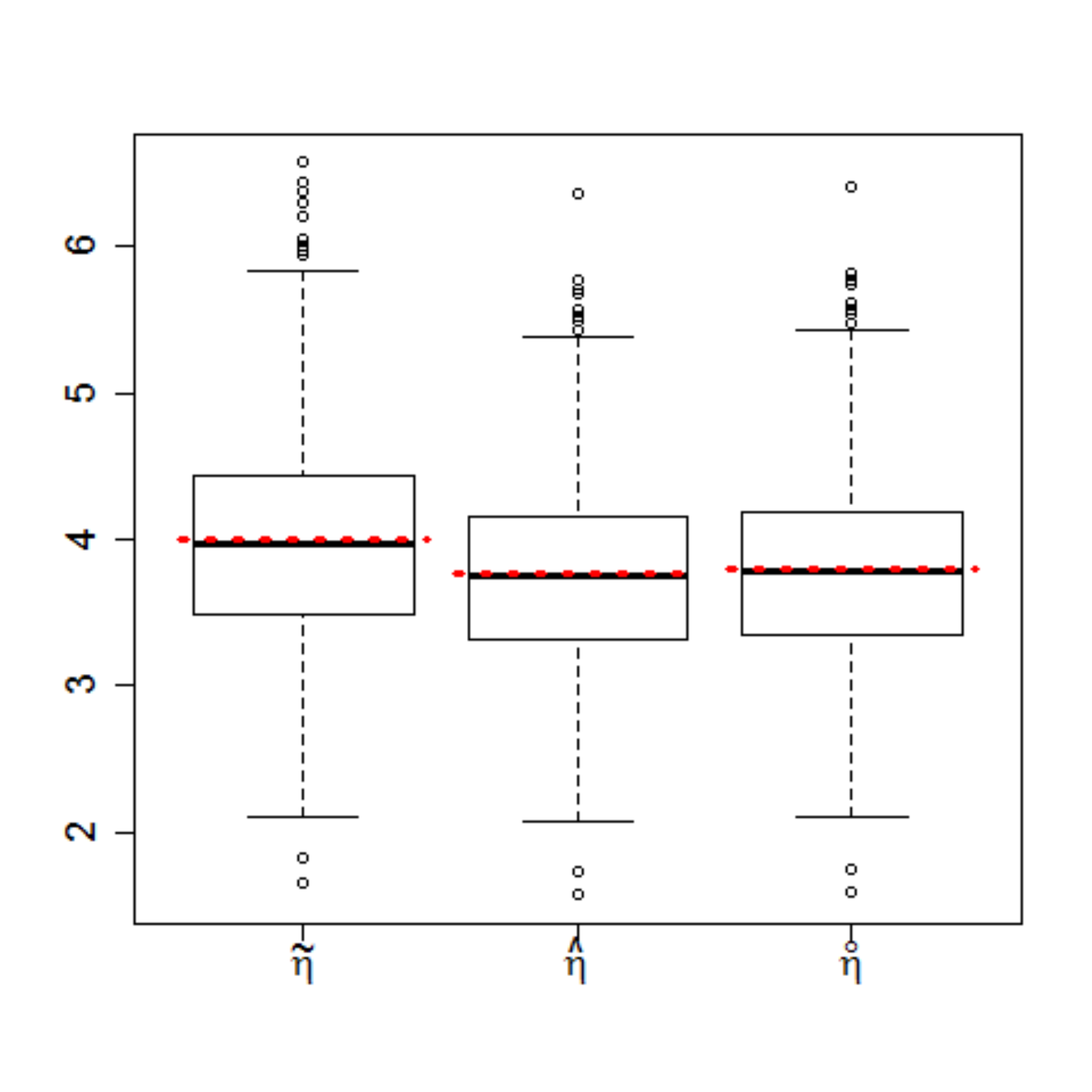}}
\scalebox{.8}{\includegraphics[width=150pt,height=200pt]{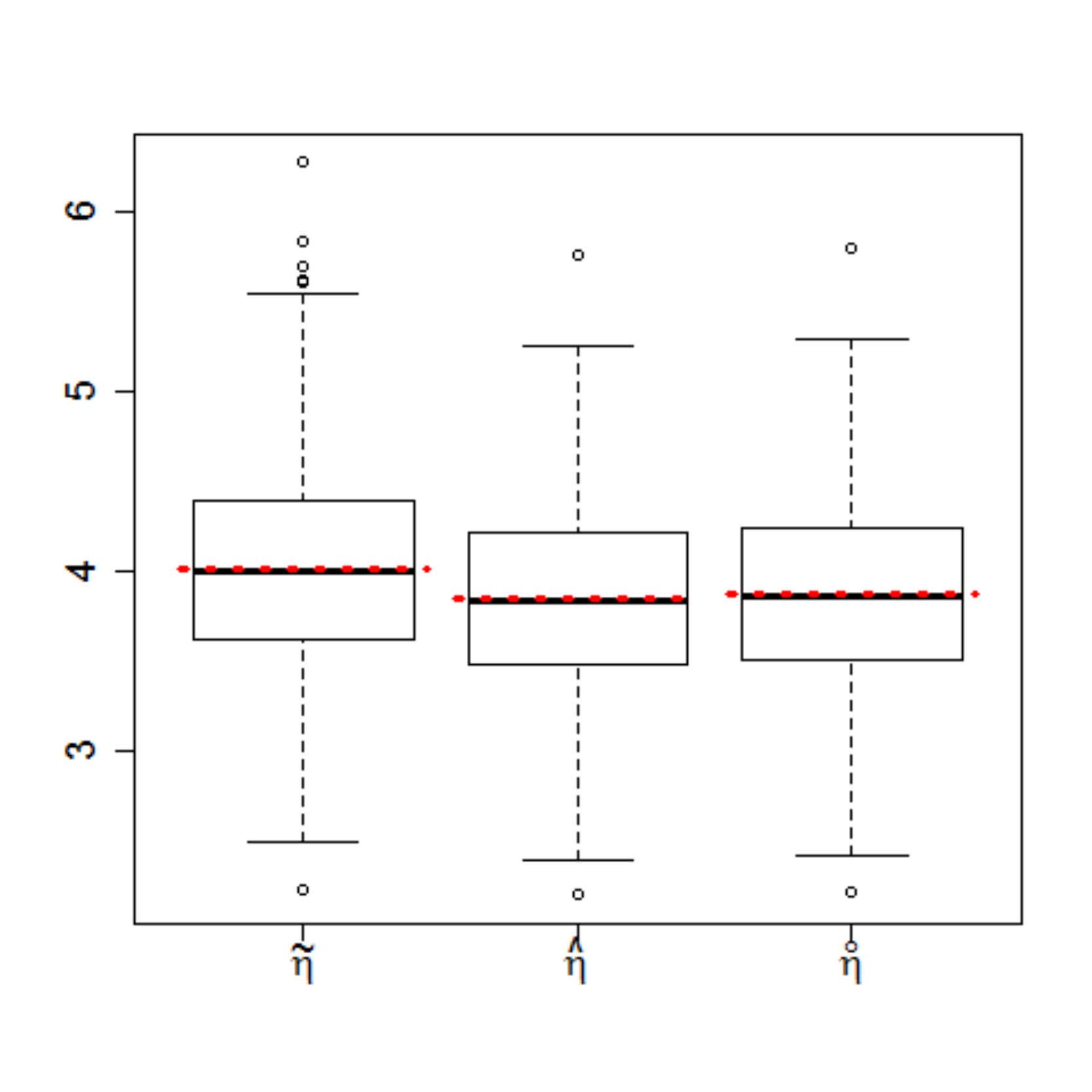}}\vspace{-2ex}\\
{\small Figure 5. Box plots for the estimator $\mathring{\eta}$ for sample sizes $n=10,20,30$ (above)\\ and for the estimators
$\tilde{\eta}$, $\hat{\eta}$ y $\mathring{\eta}$ for sample sizes $n=10,20,30$, respectively (below).}
\end{center}
Notice that both $\hat{\eta}$ and $\tilde{\eta}$ are equivariant estimators of the scale parameter $\eta$. So they have greater risk for the loss function $W_1$ than $\mathring{\eta}$. Hence (see  Table 5 and Figure 5), in the previous simulation study, the MSE of $\mathring{\eta}$ is smaller than the MSE of $\hat{\eta}$ and $\tilde{\eta}$.  $\Box$
       \end{myrema}

\begin{myrema}\rm
Although less interesting from the perspective of real applications, for completeness we now consider
the problem of estimating the scale parameter $\eta$ when the
location parameter $\xi$ is known, say $\xi=\xi_0$. After the shift
$(y_1,\dots,y_n)\mapsto (y_1-\xi_0,\dots,y_n-\xi_0)$, the
statistical model remains invariant under the transformations
(dilations) of the form $(y_1,\dots,y_n)\mapsto (ay_1,\dots,ay_n)$,
for $a>0$. For the loss function
$W'_1(\eta,x)=(x-\nolinebreak\eta)^2/\eta^2$, the MRE estimator of
the scale parameter $\eta$ is
$$T_2=\frac{\Gamma(\frac{n+1}{2})}{\sqrt{2}\Gamma(\frac{n+2}{2})}\sqrt{\sum_{i=1}^n(Y_i-\xi_0)^2}=
\frac{B(\frac{n+1}{2},\frac12)}{\sqrt{2\pi}}
\sqrt{\sum_{i=1}^n(Y_i-\xi_0)^2},
$$
where $B$ denotes Euler's beta function. In
fact, for the loss function $W'_1$, the MRE estimator of $\eta$ is
$$T_2(y_1,\dots,y_n)=\frac{\displaystyle\int_0^{\infty}v^nh_1(v(y_1-\xi_0),...,v(y_n-\xi_0))dv}
{\displaystyle\int_0^{\infty}v^{n+1}h_1(v(y_1-\xi_0),...,v(y_n-\xi_0))dv},$$
where
$$h_1(y_1,\dots,y_n)=\left(\frac{2}{\pi}\right)^{\frac{n}{2}}\exp
\left\{-\frac{1}{2}\sum_{i=1}^ny_i^2\right\}
I_{[0,+\infty[}(y_{1:n}).$$

To simplify the notation, we assume without loss of generality that
$\xi_0=0$. The change of variable
$t=\frac{1}{2}\sum_{i=1}^ny_i^2v^2$ leads to, for $k=n,n+1$,

\begin{gather*}\int_0^{\infty}v^kh_1(vy_1,...,vy_n)dv=
2^{\frac{n+k-1}{2}}\pi^{-\frac{n}{2}}\left(\sum_{i=1}^ny_i^2\right)^{-\frac{k+1}{2}}
\Gamma\left(\frac{k+1}{2}\right)I_{[0,+\infty[}(y_{1:n}),
\end{gather*}
and the assertion then follows easily.

Note also that, when $\xi=\xi_0$,
$$\frac1n\sum_{i=1}^n(Y_i-\xi_0)^2$$ is
the minimum variance unbiased estimator of $\eta^2$. This is a
consequence of the Lehmann-Scheff\'{e} Theorem and the facts that
$\sum_{i=1}^n(Y_i-\xi_0)^2$ is a sufficient and complete statistic
and $\eta^{-2}\sum_{i=1}^n(Y_i-\xi_0)^2$ has a $\chi^2(n)$ distribution. A little more work shows that
$$\frac{\Gamma(\frac{n}{2})}{\sqrt{2}\Gamma(\frac{n+1}{2})}\sqrt{\sum_{i=1}^n(Y_i-\xi_0)^2}=
\frac{B(\frac{n}{2},\frac12)}{\sqrt{2\pi}}
\sqrt{\sum_{i=1}^n(Y_i-\xi_0)^2}
$$
is the minimum variance unbiased estimator of $\eta$. $\Box$
\end{myrema}

\begin{myack}\rm
This work was supported by the Spanish Ministerio de Ciencia y
Tecnolog\'ia under the project MTM2010-16845 and the Junta de
Extremadura under the GR10064 grant.
\end{myack}


\end{document}